\documentclass[%
 reprint,
 amsmath,amssymb,
 aps,prd,
]{revtex4-2}

\usepackage{breqn}
\usepackage{amsmath}
\usepackage{color}
\usepackage{graphicx}
\usepackage{dcolumn}
\usepackage{bm}
\usepackage{hyperref}
\usepackage{booktabs}

\hypersetup{
    colorlinks=true,
    linkcolor=blue,
    citecolor=blue,
    filecolor=magenta,      
    urlcolor=blue,
    pdftitle={Constraining the Progenitor Orbits of Sequential Black-Hole Mergers in Hierarchical Triples: Peters-Based Estimates},
    pdfauthor={Felix M. Heinze, Zhoujian Cao},
    pdfpagemode=FullScreen,
}

\begin{document}

\preprint{APS/123-QED}

\title{Constraining the Progenitor Orbits of Sequential Black-Hole Mergers in \texorpdfstring{\\}{ } Hierarchical Triples: Peters-Based Estimates}

\author{Felix M. Heinze$^{1}$}
\author{Zhoujian Cao$^{2,3,4}$}
\affiliation{$^{1}$Friedrich-Schiller-Universität Jena$,$ Theoretisch Physikalisches Institut$,$ 07743 Jena$,$ Germany}
\affiliation{$^{2}$Department of Astronomy$,$ Beijing Normal University$,$ Beijing 100875$,$ China}
\affiliation{$^{3}$Institute for Frontiers in Astronomy and Astrophysics$,$ Beijing Normal University$,$ Beijing 102206$,$ China}
\affiliation{$^{4}$School of Fundamental Physics and Mathematical Sciences$,$ Hangzhou Institute for Advanced Study$,$ UCAS$,$ Hangzhou 310024$,$ China}

\begin{abstract}
Two black-hole merger events may originate from sequential mergers in a pre-existing hierarchical black-hole triple if their spatial localizations overlap and the remnant mass of the first merger is compatible with a component mass of the second. In this scenario, the inner binary merges first, leaving a remnant that remains bound to the tertiary and subsequently undergoes a second merger. We present a method to constrain the progenitor triple’s outer orbit by modeling the gravitational-wave-driven evolution of the remnant-tertiary binary using Peters’ equations. Given an assumed upper bound on the recoil velocity of the first merger, the method requires only the black-hole masses, the time delay between the mergers, and the eccentricity at a reference frequency—quantities that can be inferred from the merger events themselves. We investigate the potential constraining power by constructing a constraint atlas spanning a range of system properties and identify the conditions under which the method places tight and reliable bounds on the progenitor triple’s outer orbital parameters. By imposing an upper bound on the detector-band eccentricity, we also obtain meaningful constraints for candidate hierarchical mergers proposed in the literature, including GW190514--GW190521. We provide numerical tools for applying the method to user-specified systems and evaluating the reliability of the resulting Peters-based constraints. This framework offers a computationally inexpensive first step toward reconstructing hierarchical-merger progenitors and provides a foundation for extensions incorporating more sophisticated relativistic treatments.
\end{abstract}

                              
\maketitle


\section{\label{sec:introduction}Introduction}
Hierarchical black-hole triples, consisting of an inner binary orbiting a more distant tertiary, can form through the evolution of massive stellar triples in the field or through dynamical encounters in dense stellar environments \citep{Antognini2015, Antonini2016, Antonini2017, Liu2018, Martinez2020}. The gravitational perturbation from the tertiary can excite the eccentricity of the inner binary through Lidov--Kozai oscillations or non-secular three-body interactions, thereby accelerating its merger through gravitational-wave emission \citep{Lidov1962, Kozai1962, Naoz2016, Kimpson2016, Silsbee2017}. If the remnant of the inner-binary merger remains bound to the tertiary, the resulting remnant--tertiary binary may subsequently undergo a second merger, producing two physically related gravitational-wave events in a successive-merger chain. 

Several variants of this scenario have been explored in the literature. Ref.~\citep{Lu2020} proposed that GW190814 could have originated from a hierarchical triple in which a binary-neutron-star merger produced the secondary compact object, which subsequently merged with a black-hole tertiary. Successive compact-object mergers have also been investigated in more general hierarchical multiple systems \citep{Liu2020,Hamers2020}, including possible connections to gravitational-wave events that have been discussed as hierarchical-merger candidates, while \citep{Samsing2017} studied double gravitational-wave mergers arising from binary--single encounters. Despite differences in their dynamical origin, these successive-merger channels provide pathways to second-generation compact objects. Their comparatively high masses and remnant spins of order $\chi \sim 0.7$ can leave distinctive signatures in the mass and spin distributions of subsequent mergers \citep{Gerosa2021, Tagawa2021}. Dynamical environments capable of producing repeated mergers can also give rise to binaries with measurable eccentricity \citep{Rodriguez2018}. The gravitational-wave observation of both mergers allows a proposed association to be tested directly through the consistency of the first-merger remnant with a component of the second binary, together with their inferred sky locations, luminosity distances, and temporal ordering. Searches for such merger families have identified several candidate associations, including GW190514$\to$GW190521, GW190708$\to$GW190915, and GW190512$\to$GW190910, although none has yet been conclusively established \citep{Veske2020,Veske2021}. Recently, a multimessenger analysis reported additional Bayesian support for an association between GW190514 and GW190521 inside the active galactic nucleus J124942.3+344929, including the candidate optical counterpart ZTF19abanrhr \citep{Graham2020, Li2025}.

The broad demographics of stellar triples are observationally constrained, but their compact-object descendants remain poorly understood. Massive stars, the progenitors of stellar-mass black holes, exhibit a particularly high multiplicity, with O stars frequently belonging to triples or higher-order multiple systems. Observations of these systems reveal outer tertiaries spanning broad distributions of mass ratios and separations \citep{Sana2014, Moe2017}. However, these observed stellar distributions cannot be translated directly into priors for black-hole triples. Stellar evolution, mass transfer, mass loss, and natal kicks may substantially modify or disrupt primordial systems, while triples in dense stellar environments can instead be assembled dynamically through binary-mediated encounters \citep{Antonini2017, Martinez2020, Fragione2020}. Subsequent gravitational-wave emission can further reshape the orbital configuration of the compact-object system. Predictions for the abundance and orbital properties of black-hole triples therefore depend strongly on their assumed formation environment and evolutionary history. The population capable of producing two successive mergers is subject to even greater uncertainty, since the first-merger remnant must remain bound to the tertiary despite mass loss through gravitational radiation and the associated recoil. Future gravitational-wave observations may be able to constrain the outer orbits of hierarchical triples through Doppler shifts and variations in the arrival time of the inner binary's signal induced by the tertiary \citep{Meiron2017,Wong2019}. However, the feasibility of such methods and the attainable precision depend significantly on the system parameters and viewing angle \citep{Robson2018}. This motivates a complementary event-based approach in which the progenitor orbit is constrained using the gravitational-wave-inferred properties of a candidate pair of sequential mergers.

Previous studies of sequential-merger candidates have focused primarily on establishing whether two gravitational-wave events could be physically associated. Under the hypothesis that such an association reflects a common hierarchical-triple origin, the event pair also contains information about the dynamics of the progenitor triple. The masses inferred from the two signals constrain the mass configuration of the system, while the time between the events and the detector-band eccentricity of the second merger constrain the post-first-merger orbit of the remnant--tertiary binary and its subsequent gravitational-wave-driven evolution. Relating these constraints to the progenitor outer orbit requires accounting for the changes caused by mass loss and gravitational recoil during the first merger. The resulting inverse problem is highly degenerate because the recoil magnitude and direction, as well as the position of the inner binary with respect to the tertiary at the time of the first merger, are generally unknown. Moreover, post-merger binaries with only a few remaining orbital periods or very small pericenter separations may lie outside the regime in which orbit-averaged evolution based on the Peters equations is reliable \citep{Peters1963, Peters1964, Zwick2020, Fumagalli2025}. A systematic framework is therefore needed both to map the measured event properties onto possible progenitor orbits and to determine when the resulting inference can be trusted.

In this study, we address both aspects by developing an event-based method for placing constraints on the progenitor outer-orbit parameters of a candidate hierarchical-triple merger and assessing their reliability. Our method relates the inferred properties of a candidate merger pair to the outer orbit of its proposed progenitor triple. For each possible outer-orbit configuration, we determine the minimum recoil velocity required at the first merger to produce a remnant–tertiary binary whose Peters evolution matches the observed inter-merger delay and the adopted eccentricity constraint for the second merger. An assumed upper bound on the recoil velocity then restricts the allowed progenitor configurations, yielding constraints on the outer semi-major axis, eccentricity, and orbital phase at the first merger. We additionally assess the reliability of the Peters approximation using criteria based on the number of remaining orbits and the pericenter separation, and compare representative systems with direct post-Newtonian integrations. To systematically investigate how these constraints and their reliability depend on the properties inferred from the event pair, we construct a constraint atlas spanning the masses and detector-band eccentricity of the remnant-tertiary binary, as well as the inter-merger delay. The numerical tools developed in this work are made available for application to other candidate systems. \\

The paper is structured as follows. In Sec.~\ref{sec:physical_setup}, we introduce the physical setup and inverse problem in greater detail. Sec.~\ref{sec:peters_constraints} describes the Peters-based inference of the progenitor outer-orbit parameters using minimum-recoil-velocity maps. We additionally introduce measures for quantifying the strength of the resulting constraints and describe the construction of a constraint atlas used to investigate the dependence of the constraints on the system properties. In Sec.~\ref{sec:peters_validity}, we assess the applicability of the Peters formalism to the problem at hand and validate representative results using post-Newtonian simulations. Sec.~\ref{sec:results} is dedicated to the main results of our study. These include the full constraint atlas as well as constraints on the outer orbits of potential hierarchical-merger candidates identified in the literature. Finally, we discuss the implications and limitations of our results and summarize our conclusions in Sec.~\ref{sec:summary}.

\begin{figure*}[htp!]
    \centering
    \includegraphics[width=1\textwidth]{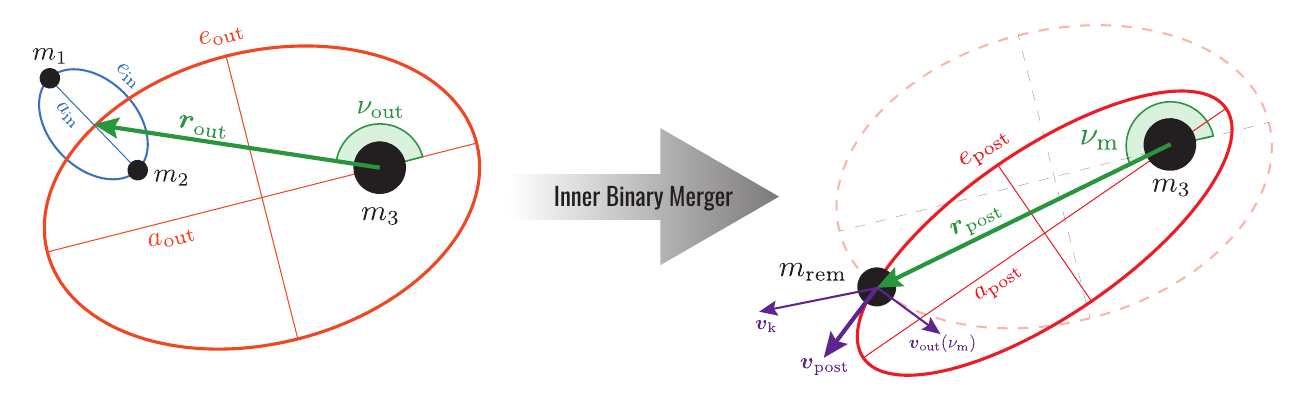}
    \caption{Left: Hierarchical triple setup before the first merger. The inner and outer orbits, together with their corresponding orbital parameters, are shown in blue and orange, respectively. Right: Configuration immediately after the first merger. The inner binary has formed a remnant black hole of mass $m_{\rm rem}$, which receives a recoil velocity $\boldsymbol{v}_k$. The recoil changes the orbital motion relative to the tertiary and places the remnant–tertiary binary on a new post-first-merger orbit (shown in red), from which the second merger may subsequently occur. The quantities indicated in the figure are defined in the main text.}
    \label{fig:double_merger_concept}
\end{figure*}

\section{Physical setup and inverse problem}
\label{sec:physical_setup}

We consider a hierarchical black-hole triple immediately before the first of two successive mergers. The system consists of an inner binary with component masses $m_1$ and $m_2$, whose center of mass and a tertiary black hole of mass $m_3$ form a wider outer binary. The hierarchical nature of the system is characterized by well-separated orbital scales, such that the spatial extent of the inner binary remains small compared to the separation from the tertiary. The inner binary merges first, producing a remnant of mass $m_{\rm rem}$ that receives a recoil kick $\boldsymbol{v}_{\rm k}$ due to the anisotropic emission of gravitational waves. If the remnant remains bound to the tertiary, the resulting remnant--tertiary binary may subsequently merge and produce a second gravitational-wave event. The setup is depicted schematically in Fig.~\ref{fig:double_merger_concept}. Our objective is not to reconstruct the full evolution that led the inner binary to its first merger, but to determine which outer-orbit configurations immediately before that merger are compatible with the inferred properties of the two events.

\subsection{Hierarchical-triple configuration}
\label{sec:triple_configuration}

We denote the total mass of the inner binary before the first merger by
\begin{equation}
    m_{12} = m_1 + m_2,
\end{equation}
and the total mass of the triple at this time by
\begin{equation}
    M_{\rm pre} = m_{12} + m_3.
\end{equation}
The Keplerian elements of the inner and outer orbits can be generally written as
\begin{equation}
    \mathcal{O}_{\alpha}
    =
    \left(
        a_{\alpha},
        e_{\alpha},
        i_{\alpha},
        \Omega_{\alpha},
        \omega_{\alpha},
        \nu_{\alpha}
    \right),
    \qquad
    \alpha\in\{\mathrm{in},\mathrm{out}\},
    \label{eq:general_orbital_elements}
\end{equation} where $a_{\alpha}$ is the semi-major axis, $e_{\alpha}$ is the eccentricity, $i_{\alpha}$ is the inclination relative to a chosen reference plane, $\Omega_{\alpha}$ is the longitude of the ascending node, $\omega_{\alpha}$ is the argument of pericenter, and $\nu_{\alpha}$ is the true anomaly. 

In the following, the motion of the inner binary's center of mass relative to the tertiary immediately before the first merger is approximated as a Newtonian two-body orbit, treating the inner binary as a point mass equal to its total mass. This treatment neglects relativistic corrections to the outer motion as well as perturbations arising from the finite size and internal dynamics of the inner binary. These approximations are expected to be adequate when the system is sufficiently hierarchical and the outer orbit remains in the weak-field regime.

The absolute orientation of the outer orbit relative to an external reference frame does not affect the intrinsic dynamics considered here. We therefore place the coordinate origin at the tertiary and choose the $x$--$y$ plane to coincide with the outer orbital plane, with the positive $x$-axis pointing toward pericenter and the positive $z$-axis aligned with the outer orbital angular momentum. The positive $y$-axis completes a right-handed coordinate system.

\noindent
We define the radial and transverse unit vectors as
\begin{align}
    \hat{\boldsymbol r}
    &=
    \cos \nu_{\rm out}\,\hat{\boldsymbol x}
    +
    \sin \nu_{\rm out}\,\hat{\boldsymbol y},
    \label{eq:outer_radial_unit_vector}
    \\
    \hat{\boldsymbol \phi}
    &=
    -\sin \nu_{\rm out}\,\hat{\boldsymbol x}
    +
    \cos \nu_{\rm out}\,\hat{\boldsymbol y}.
    \label{eq:outer_tangential_unit_vector}
\end{align}
and the unit vector normal to the outer orbital plane as
\begin{equation}
    \hat{\boldsymbol n}
    =
    \hat{\boldsymbol r}
    \times
    \hat{\boldsymbol \phi}
    =
    \hat{\boldsymbol z}.
    \label{eq:outer_normal_unit_vector}
\end{equation}
The relative position of the inner binary's center of mass with respect to the tertiary is therefore given by
\begin{equation}
    \boldsymbol{r}_{\rm out}
    =
    r_{\rm out}\hat{\boldsymbol r}.
    \label{eq:outer_position}
\end{equation}
The relative separation is given by
\begin{equation}
    r_{\rm out}
    =
    \frac{p_{\rm out}}
    {1+e_{\rm out}\cos \nu_{\rm out}},
    \label{eq:outer_separation}
\end{equation}
and the radial and transverse components of the relative velocity are
\begin{align}
    v_{r, \rm out}
    =
    \dot r_{\rm out}
    &=
    \sqrt{\frac{G M_{\rm pre}}{p_{\rm out}}}\,
    e_{\rm out}\sin \nu_{\rm out},
    \label{eq:outer_radial_velocity}
    \\
    v_{\phi, \rm out}
    =
    r_{\rm out}\dot \nu_{\rm out}
    &=
    \sqrt{\frac{G M_{\rm pre}}{p_{\rm out}}}\,
    \left(1+e_{\rm out}\cos \nu_{\rm out}\right),
    \label{eq:outer_transverse_velocity}
\end{align}
where we introduced the semilatus rectum
\begin{equation}
    p_{\rm out}
    =
    a_{\rm out}\left(1-e_{\rm out}^{2}\right).
    \label{eq:outer_semilatus_rectum}
\end{equation}
The relative velocity vector is therefore given by
\begin{equation}
    \boldsymbol{v}_{\rm out}
    = 
    v_{r, \rm out}\hat{\boldsymbol r}
    +
    v_{\phi, \rm out}\hat{\boldsymbol \phi}.
    \label{eq:outer_velocity}
\end{equation}
Consequently, the three quantities $\left(a_{\rm out}, e_{\rm out}, \nu_{\rm out} \right)$ determine the instantaneous outer-orbit phase-space state $(\boldsymbol{r}_{\rm out},\boldsymbol{v}_{\rm out})$ once the component masses are specified.

\subsection{The first merger and the remnant--tertiary binary}
\label{sec:first_merger_mapping}

We model the first merger as an instantaneous event. Accordingly, the outer separation is unchanged during the merger, while the remnant mass and relative velocity change discontinuously due to gravitational-wave mass loss and recoil. The first-merger remnant has mass
\begin{equation}
    m_{\rm rem}
    =
    m_{12}-\Delta m
    =
    \left(1-\epsilon_{\rm rad}\right)m_{12},
    \label{eq:remnant_mass}
\end{equation}
where $\Delta m$ is the mass-energy radiated in gravitational waves and $\epsilon_{\rm rad}$ is the corresponding fractional mass loss. The remnant additionally receives a recoil velocity
\begin{equation}
    \boldsymbol{v}_{\rm k}
    =
    v_{{\rm k},r}\hat{\boldsymbol{r}}
    +
    v_{{\rm k},\phi}\hat{\boldsymbol{\phi}}
    +
    v_{{\rm k},n}\hat{\boldsymbol{n}},
    \label{eq:kick_decomposition}
\end{equation}
where the components are defined with respect to the radial, transverse, and normal directions of the outer orbit at the time of the first merger.

Immediately after the first merger, the relative position and velocity of the remnant with respect to the tertiary are therefore given by
\begin{align}
    \boldsymbol{r}_{\rm post}
    &=
    \boldsymbol{r}_{\rm out}(\nu_{\rm m}),
    \label{eq:post_position}
    \\
    \boldsymbol{v}_{\rm post}
    &=
    \boldsymbol{v}_{\rm out}(\nu_{\rm m})
    +
    \boldsymbol{v}_{\rm k},
    \label{eq:post_velocity}
\end{align}
where $\nu_{\rm m}$ is the outer orbit's true anomaly at the time of the first merger. The total mass of the remnant--tertiary binary is
\begin{equation}
    M = m_{\rm rem}+m_3,
    \label{eq:second_binary_total_mass}
\end{equation}
and its symmetric mass ratio is
\begin{equation}
    \eta
    =
    \frac{m_{\rm rem}m_3}{M^2}.
    \label{eq:second_binary_eta}
\end{equation}
The specific orbital energy and specific angular-momentum vector of the post-first-merger binary are
\begin{align}
    \varepsilon_{\rm post}
    &=
    \frac{1}{2}
    \left|\boldsymbol{v}_{\rm post}\right|^2
    -
    \frac{GM}{r_{\rm post}},
    \label{eq:post_energy}
    \\
    \boldsymbol{h}_{\rm post}
    &=
    \boldsymbol{r}_{\rm post}
    \times
    \boldsymbol{v}_{\rm post}.
    \label{eq:post_angular_momentum}
\end{align}
A bound remnant--tertiary binary is formed only if
\begin{equation}
    \varepsilon_{\rm post}<0.
    \label{eq:bound_condition}
\end{equation}
For a bound configuration, its osculating semi-major axis is given by
\begin{equation}
    a_{\rm post}
    =
    -\frac{GM}{2\varepsilon_{\rm post}},
    \label{eq:post_semi-major_axis}
\end{equation}
while its eccentricity vector and eccentricity are
\begin{align}
    \boldsymbol{e}_{\rm post}
    &=
    \frac{
        \boldsymbol{v}_{\rm post}
        \times
        \boldsymbol{h}_{\rm post}
    }{GM}
    -
    \frac{\boldsymbol{r}_{\rm post}}{r_{\rm post}},
    \label{eq:post_eccentricity_vector}
    \\
    e_{\rm post}
    &=
    \left|\boldsymbol{e}_{\rm post}\right|.
    \label{eq:post_eccentricity}
\end{align}
Equations~\eqref{eq:outer_position}--\eqref{eq:post_eccentricity}
therefore provide the instantaneous mapping
\begin{equation}
    \left(
        a_{\rm out}, e_{\rm out}, \nu_{\rm m},
        \boldsymbol{v}_{\rm k};
        m_{12},
        m_3,
        m_{\rm rem}
    \right)
    \longrightarrow
    \left(a_{\rm post},e_{\rm post}\right).
    \label{eq:instantaneous_mapping}
\end{equation}

\subsection{Merger modeling and parameter-space reduction}
\label{sec:orientation_reduction}

In a general hierarchical triple, the mutual inclination and the relative orientations of the lines of nodes and apsides influence the preceding three-body evolution. They may, for example, determine whether the inner binary undergoes secular eccentricity excitation or enters a nonsecular regime. In addition, the outcome of the first merger depends on the inner-binary mass ratio and on the component spins and their orientations at merger. These properties affect the energy radiated in gravitational waves, the mass and spin of the merger remnant, and the magnitude and direction of the gravitational recoil.

The analysis presented here uses only the inferred masses, inter-merger delay, and detector-band eccentricity, without incorporating constraints on the relative orientation of the inner and outer binaries, the individual spin magnitudes and orientations, or the recoil-velocity vector. We therefore do not attempt to reconstruct these quantities or the pre-merger evolution of the inner binary. Instead, we allow the recoil-velocity vector to vary freely and restrict the reconstructed progenitor parameters to the outer eccentricity, semi-major axis, and true anomaly at the time of the first merger,
\begin{equation}
    \boldsymbol{\theta}_{\rm out}
    =
    \left(a_{\rm out},e_{\rm out},\nu_{\rm m}\right).
\end{equation}

We likewise do not model the first merger and the properties of its remnant self-consistently from the detailed inner-binary configuration at merger. Phenomenological prescriptions calibrated to numerical-relativity simulations can be used to relate the binary mass ratio and component spins to the radiated energy, remnant spin, and recoil velocity \cite{Lousto2010,Barausse2012,Healy2014,Zlochower2015}. For the present reduced construction, we instead prescribe a fixed fractional mass loss $\epsilon_{\rm rad}$ and treat the recoil vector $\boldsymbol{v}_k$ as an independent free quantity. Spin-dependent effects on the first-merger recoil are therefore effectively absorbed into the freely varying $\boldsymbol{v}_k$, without attempting to infer the underlying spin configuration.

We additionally neglect spin-orbit and spin-spin effects in the subsequent evolution of the remnant--tertiary binary. Spin-dependent contributions enter at higher post-Newtonian order and are therefore expected to be comparatively small in the weak-field, secular regime relevant to our Peters-based constraints. Their influence may become more important for configurations approaching the strong-field or few-passage regime.

Possible extensions incorporating information on the orbital orientations and black-hole spins, together with numerical-relativity-calibrated prescriptions relating the inner-binary properties to the mass, spin, and recoil of the first-merger remnant, are discussed in Sec.~\ref{sec:summary}.

\subsection{Dynamical stability and physically admissible progenitor domain}
\label{sec:physical_domain}

A clear separation of the inner and outer orbital scales does not by itself guarantee that a hierarchical triple remains dynamically stable. In particular, systems with a highly eccentric outer orbit may undergo strong interactions near outer pericenter even when $a_{\rm in}\ll a_{\rm out}$. We therefore restrict the progenitor parameter space to configurations compatible with a stability criterion for hierarchical triples.

Several analytic, empirical, and numerical methods have been developed to assess the stability of hierarchical triples. In the present work, we adopt the Mardling--Aarseth criterion because it provides a simple but reliable and widely used relation between the inner and outer orbital scales \citep{MardlingAarseth2001}. In its standard form, it can be written as
\begin{equation}
    \frac{a_{\rm out}}{a_{\rm in}}
    >
    \frac{2.8}{1-e_{\rm out}}
    \left[
        \left(1+q_{\rm out}\right)
        \frac{1+e_{\rm out}}
             {\sqrt{1-e_{\rm out}}}
    \right]^{2/5}
    \left(
        1-\frac{0.3 i}{\pi}
    \right),
    \label{eq:ma_stability_criterion}
\end{equation}
where
\begin{equation}
    q_{\rm out}
    =
    \frac{m_3}{m_1+m_2}
    =
    \frac{m_3}{m_{12}}
    \label{eq:outer_mass_ratio}
\end{equation}
is the mass ratio between the tertiary and the inner binary, and $i\in[0,\pi]$ is the mutual inclination between the inner and outer orbital planes. Here, $a_{\rm in}$ denotes the osculating semi-major axis of the inner binary at the epoch at which stability is assessed.

The mutual inclination is not reconstructed in the present analysis. Rather than imposing an arbitrary orientation, we retain an outer-orbit configuration whenever it satisfies the Mardling--Aarseth criterion for at least one possible value of $i$. Within the standard form of the criterion, the least restrictive stability boundary is obtained for $i=\pi$. This choice does not imply that the progenitor triple is assumed to be retrograde. Instead, it defines an orientation-agnostic compatibility condition: a configuration is excluded only if it fails the adopted stability criterion for every possible mutual inclination. Because this choice maximizes the admissible progenitor domain, it yields the weakest, and hence most conservative, constraints on the outer-orbit parameters.

For later convenience, Eq.~\eqref{eq:ma_stability_criterion} can be rewritten to give the smallest outer semi-major axis compatible with a given inner binary. Setting $i=\pi$ gives
\begin{equation}
    a_{\rm out,min}^{\rm MA}
    =
    1.96\,a_{\rm in}
    \left(1+q_{\rm out}\right)^{2/5}
    \frac{\left(1+e_{\rm out}\right)^{2/5}}
         {\left(1-e_{\rm out}\right)^{6/5}} .
    \label{eq:aoutmin_MA}
\end{equation}
For a specified inner semi-major axis $a_{\rm in}$, we define the physically admissible progenitor domain as
\begin{equation}
\begin{aligned}
    \mathcal D_{\rm phys}
    =
    \left\{
        \boldsymbol{\theta}_{\rm out}\in\mathcal D_{\rm ref}
        \,\middle|\,
        a_{\rm out}
        >
        a_{\rm out,min}^{\rm MA}
        (e_{\rm out};a_{\rm in},m_{12},m_3)
    \right\}.
    \label{eq:Dphys}
\end{aligned}
\end{equation}
Here, $\mathcal D_{\rm ref}$ denotes the finite reference domain over which the progenitor parameters are scanned. The choice of $a_{\rm in}$ specifies the pre-first-merger epoch at which stability is assessed. A larger $a_{\rm in}$, corresponding to an earlier stage of the inner inspiral, produces a more restrictive stability condition.

The Mardling--Aarseth relation should be regarded as an approximate admissibility criterion rather than an exact boundary between stable and unstable systems. Its standard form contains no explicit dependence on the inner eccentricity or orbital phases and includes only a simple correction for mutual inclination. More recent stability prescriptions incorporate additional dependences on these parameters or use direct numerical and machine-learning classifications \citep{Myllari2018, Vynatheya2022, Tory2022}. We nevertheless adopt the Mardling--Aarseth criterion as our fiducial prescription because of its simplicity and because it provides a reliable, transparent, and physically motivated lower boundary for the outer-orbit parameter space.

\subsection{Event-pair inputs and formulation of the inverse problem}
\label{sec:inverse_problem}

Under the sequential-merger hypothesis, the two gravitational-wave signals provide information on the masses $m_1$, $m_2$, $m_3$ , and $m_{\rm rem}$ relevant to the progenitor triple. A physical association requires the remnant mass inferred from the first event to be consistent with one of the component masses inferred from the second event.

The temporal separation between the two events, $\Delta t$, is interpreted as the lifetime of the remnant--tertiary binary. The second merger event may additionally allow for constraining its orbital eccentricity $e_{\rm ref}$ at a reference gravitational-wave frequency $f_{\rm ref}$. We allow the detector-band eccentricity information to take two forms. For a measurement with central value $e_{{\rm ref},c}$ and half-width $\delta e$, we define
\begin{equation}
    \mathcal E_{\rm ref}
    =
    \left[
        e_{{\rm ref},c}-\delta e,
        e_{{\rm ref},c}+\delta e
    \right]
    \cap[0,1),
    \label{eq:eccentricity_interval_measurement}
\end{equation}
whereas an upper limit $e_{{\rm ref},\max}$ corresponds to
\begin{equation}
    \mathcal E_{\rm ref}
    =
    \left[
        0,e_{{\rm ref},\max}
    \right].
    \label{eq:eccentricity_interval_upper}
\end{equation} 
With a fixed reference frequency $f_{\rm ref}$, we therefore characterize an individual candidate system by
\begin{equation}
    \boldsymbol{\lambda}
    =
    \left(
        m_1,
        m_2,
        m_{\rm rem},
        m_3,
        \Delta t,
        \mathcal E_{\rm ref}
    \right).
    \label{eq:general_system_parameters}
\end{equation}
Throughout, all masses, the inter-merger delay $\Delta t$, and the reference frequency $f_{\rm ref}$ refer to the detector frame, i.e.~to quantities as measured by the observer.

For each physically admissible progenitor outer-orbit configuration $
    \boldsymbol{\theta}_{\rm out}
    =
    \left(
        a_{\rm out},
        e_{\rm out},
        \nu_{\rm m}
    \right)
    \in \mathcal D_{\rm phys}$ 
we determine whether there exist recoil velocity vectors $\boldsymbol{v}_{\rm k}$ that, together with the mass loss, produce a bound remnant--tertiary binary whose subsequent evolution is consistent with the inferred inter-merger delay and detector-band eccentricity. Minimizing the recoil velocities over the accessible post-first-merger orbits yields a minimum-recoil-velocity map, whose construction is discussed in detail in Section \ref{sec:min_recoil_maps}. Conceptually, the minimum required recoil velocity for a given system $\boldsymbol{\lambda}$ at a given progenitor outer-orbit configuration $\boldsymbol{\theta}_{\rm out}$ can be written as
\begin{equation}
    v_{\rm k,min}
    \left(
        \boldsymbol{\theta}_{\rm out};
        \boldsymbol{\lambda}
    \right)
    =
    \min_{\boldsymbol{v}_{\rm k}}
    \left\{
        \left|\boldsymbol{v}_{\rm k}\right|
        \,\middle|\,
        \begin{array}{l}
        t_{\rm m}(a_{\rm post},e_{\rm post})
        =\Delta t,\\[2pt]
        e_{\rm ref}(a_{\rm post},e_{\rm post})
        \in \mathcal E_{\rm ref}
        \end{array}
    \right\}.
    \label{eq:min_kick_definition}
    \vspace{5pt}
\end{equation}
Here, $t_{\rm m}(a_{\rm post},e_{\rm post})$ denotes the merger time obtained by evolving the post-first-merger binary according to an adopted gravitational-wave-driven model, while $e_{\rm ref}(a_{\rm post},e_{\rm post})$ denotes its corresponding eccentricity when it reaches the reference frequency $f_{\rm ref}$. For each kick velocity vector $\boldsymbol{v}_{\rm k}$, the orbital elements $a_{\rm post}$ and $e_{\rm post}$ follow from the instantaneous mapping (\ref{eq:instantaneous_mapping}) described in Sec.~\ref{sec:first_merger_mapping}, while the subsequent gravitational-wave-driven evolution is modeled using the Peters equations as described in Sec.~\ref{sec:peters_targets}.

For an adopted upper recoil limit $v_{\rm k,max}$, the dynamically allowed progenitor region is given by the progenitor outer-orbit configurations for which $v_{\rm k,min}
    \left(
        \boldsymbol{\theta}_{\rm out};
        \boldsymbol{\lambda}
    \right) \leq v_{\rm k,max}$.
The use of upper limits on the recoil velocity is motivated by the absence of precise recoil measurements for current events. Physically motivated upper limits can nevertheless be adopted from numerical-relativity predictions for the relevant black-hole population. If future observations independently constrain the recoil magnitude or direction, the inverse problem may instead be formulated by requiring consistency with the corresponding measured interval or posterior.

The formulation presented here retains the full mass information available for an individual candidate event pair. A reduced mass parametrization used to construct the constraint atlas is introduced separately in Sec.~\ref{sec:constraint_atlas}.

\section{Peters-based constraint construction}
\label{sec:peters_constraints}

We now describe the formalism used to solve the inverse problem introduced in Sec.~\ref{sec:inverse_problem}. For fixed component masses, the inter-merger delay and detector-band eccentricity define a set of possible remnant--tertiary orbits immediately after the first merger. We refer to these as the post-first-merger target orbits. For each progenitor outer-orbit configuration, and for the prescribed mass loss, we determine the recoil velocity required to reach each kinematically accessible target orbit. Minimizing over the accessible targets then yields the minimum recoil velocity required for that progenitor configuration to reproduce the prescribed event-pair properties. Evaluating this quantity throughout the progenitor parameter space produces the minimum-recoil-velocity maps on which the subsequent constraints are based.

Throughout this section, the subsequent evolution of the remnant--tertiary binary is modeled using the orbit-averaged Peters equations. Their regime of applicability is assessed separately in Sec.~\ref{sec:peters_validity}.

\subsection{Peters evolution and post-first-merger target orbits}
\label{sec:peters_targets}

Following the first merger, the remnant of mass $m_{\rm rem}$ and the tertiary of mass $m_3$ form a binary with total mass $M = m_{\rm rem}+m_3$ and initial orbital elements $(a_{\rm post},e_{\rm post})$. Within the Peters approximation, their orbit-averaged evolution is governed by the Peters equations \citep{Peters1964},
\begin{align}
    \frac{{\rm d}a}{{\rm d}t}
    &=
    -\frac{64}{5}
    \frac{G^3m_{\rm rem}m_3M}
         {c^5a^3(1-e^2)^{7/2}}
    \left(
        1+\frac{73}{24}e^2+\frac{37}{96}e^4
    \right),
    \label{eq:peters_da}
    \\
    \frac{{\rm d}e}{{\rm d}t}
    &=
    -\frac{304}{15}
    \frac{G^3m_{\rm rem}m_3M}
         {c^5a^4(1-e^2)^{5/2}}
    e
    \left(
        1+\frac{121}{304}e^2
    \right).
    \label{eq:peters_de}
\end{align}
For $e>0$, combining these equations leads to the conserved quantity
\begin{equation}
    c_0
    =
    a(1-e^2)e^{-12/19}
    \left(
        1+\frac{121}{304}e^2
    \right)^{-870/2299},
    \label{eq:peters_c0}
\end{equation}
and therefore
\begin{equation}
    a(e)
    =
    c_0
    \frac{e^{12/19}}{1-e^2}
    \left(
        1+\frac{121}{304}e^2
    \right)^{870/2299}.
    \label{eq:peters_a_e}
\end{equation}
Defining
\begin{equation}
    \mathcal I(e)
    =
    \int_0^e
    \frac{
        {e'}^{29/19}
        \left(
            1+\frac{121}{304}{e'}^2
        \right)^{1181/2299}
    }{
        (1-{e'}^2)^{3/2}
    }
    \,{\rm d}e',
    \label{eq:peters_integral}
\end{equation}
the merger time of a post-first-merger orbit $(a_{\rm post},e_{\rm post})$ is given by
\begin{equation}
    t_{\rm m}(a_{\rm post},e_{\rm post})
    =
    \frac{12}{19}
    \frac{c_0^4}{\beta}
    \mathcal I(e_{\rm post}),
    \label{eq:peters_merger_time}
\end{equation}
where
\begin{equation}
    \beta
    =
    \frac{64}{5}
    \frac{G^3m_{\rm rem}m_3M}{c^5},
    \label{eq:peters_beta}
\end{equation}
and where $c_0$ can be evaluated from $(a_{\rm post},e_{\rm post})$ using Eq.~\eqref{eq:peters_c0}.

For fixed component masses and given inter-merger delay $\Delta t$ (and assuming that $t_{\rm m} = \Delta t$), Eq.~\eqref{eq:peters_merger_time} can be inverted explicitly for the post-first-merger semi-major axis,
\begin{equation}
    a_{\rm post}
    =
    \left[
        \frac{19\beta\Delta t}
        {12\,\mathcal I(e_{\rm post})}
    \right]^{\frac{1}{4}}
    \frac{
        e_{\rm post}^{12/19}
    }{
        1-e_{\rm post}^2
    }
    \left(
        1+\frac{121}{304}e_{\rm post}^2
    \right)^{\frac{870}{2299}}.
    \label{eq:a_from_delay}
\end{equation}
The above equations are valid for $e>0$. In the circular limit one obtains
\begin{equation}
    t_{\rm m}(a_{\rm post},e_{\rm post}=0)
    =
    \frac{a_{\rm post}^4}{4\beta},
    \label{eq:peters_merger_time_circular}
\end{equation}
and hence
\begin{equation}
    a_{\rm post}(e_{\rm post}=0)
    =
    \left(4\beta\Delta t\right)^{1/4}.
    \label{eq:a_from_delay_circ}
\end{equation}
Thus, for fixed component masses and $\Delta t$, every choice of $e_{\rm post}$ defines a unique $a_{\rm post}(e_{\rm post})$ such that the Peters merger time of the corresponding post-first-merger orbit $(a_{\rm post},e_{\rm post})$ matches the inter-merger delay.

To relate these post-first-merger orbits to eccentricity information in the detector band, we adopt the commonly used approximation of \citet{Wen2003} for the characteristic peak gravitational-wave frequency of an eccentric binary,
\begin{equation}
    f_{\rm peak}(a,e)
    =
    \frac{1}
    {\pi} \sqrt{\frac{GM}{a^3(1-e^2)^3}}
    (1+e)^{1.1954}.
    \label{eq:wen_peak_frequency}
\end{equation}
For $e=0$, this expression reduces to twice the orbital frequency, as expected for the dominant quadrupolar harmonic of a circular binary. Using Eq.~\eqref{eq:peters_a_e}, the peak frequency along a Peters trajectory for $e>0$ can equivalently be written as
\begin{equation}
    f_{\rm peak}(e)
    =
    \frac{\sqrt{GM}}
    {\pi c_0^{3/2}}
    e^{-\frac{18}{19}}
    \left(
        1+\frac{121}{304}e^2
    \right)^{-\frac{1305}{2299}}
    (1+e)^{1.1954}.
    \label{eq:wen_frequency_along_peters}
\end{equation}
If the binary is initially below the detector reference frequency,
\begin{equation}
    f_{\rm peak}
    \left(
        a_{\rm post}(e_{\rm post}),e_{\rm post}
    \right)
    <
    f_{\rm ref},
    \label{eq:initial_below_band}
\end{equation}
the corresponding eccentricity $e_{\rm ref}$ is defined by
\begin{equation}
    f_{\rm peak}
    \left(
        a(e_{\rm ref}),e_{\rm ref}
    \right)
    =
    f_{\rm ref},
    \qquad
    0\leq e_{\rm ref}\leq e_{\rm post}.
    \label{eq:detector_frequency_condition}
\end{equation}
For $e_{\rm ref}>0$, using Eq.~\eqref{eq:wen_frequency_along_peters}, this can also be written as
\begin{equation}
    \frac{\sqrt{GM}}
    {\pi c_0^{3/2}}
    e_{\rm ref}^{-\frac{18}{19}}
    \left(
        1+\frac{121}{304}e_{\rm ref}^2
    \right)^{-\frac{1305}{2299}}
    (1+e_{\rm ref})^{1.1954} = f_{\rm ref}.
    \label{eq:detector_eccentricity_root}
\end{equation}
We solve this equation numerically for $e_{\rm ref}$, with $c_0$ determined from $e_{\rm post}$ and the corresponding $a_{\rm post}(e_{\rm post})$. Since $f_{\rm peak}(e)$ is monotonic, the solution for $e_{\rm ref}$ is unique. For fixed masses, $\Delta t$, and $f_{\rm ref}$, this defines the mapping $e_{\rm ref} = e_{\rm ref}(a_{\rm post}(e_{\rm post}), e_{\rm post})$ for binaries initially below the detector reference frequency. For an initially circular post-first-merger orbit, the Peters evolution remains circular, and hence $e_{\rm ref} = 0$.

If the post-first-merger binary already satisfies
\begin{equation}
    f_{\rm peak}
    \left(
        a_{\rm post}(e_{\rm post}),e_{\rm post}
    \right)
    \geq
    f_{\rm ref},
    \label{eq:initial_already_in_band}
\end{equation}
we assign $e_{\rm ref}=e_{\rm post}$ for the purposes of the target classification and flag the corresponding target as already being in the detector band.

We define the set of post-first-merger target orbits whose Peters evolution is compatible with the observed inter-merger delay and detector-band eccentricity information as
\begin{equation}
    \mathcal T_{\rm obs}(\boldsymbol{\lambda})
    =
    \left\{
        (a_{\rm post},e_{\rm post})
        \,\middle|\,
        \begin{array}{l}
        t_{\rm m}(a_{\rm post},e_{\rm post})
        =\Delta t,\\[2pt]
        e_{\rm ref}(a_{\rm post},e_{\rm post})
        \in\mathcal E_{\rm ref}
        \end{array}
    \right\}.
    \label{eq:target_post_orbits}
\end{equation}

\begin{figure*}[htp!]
    \centering
    \includegraphics[width=1\textwidth]{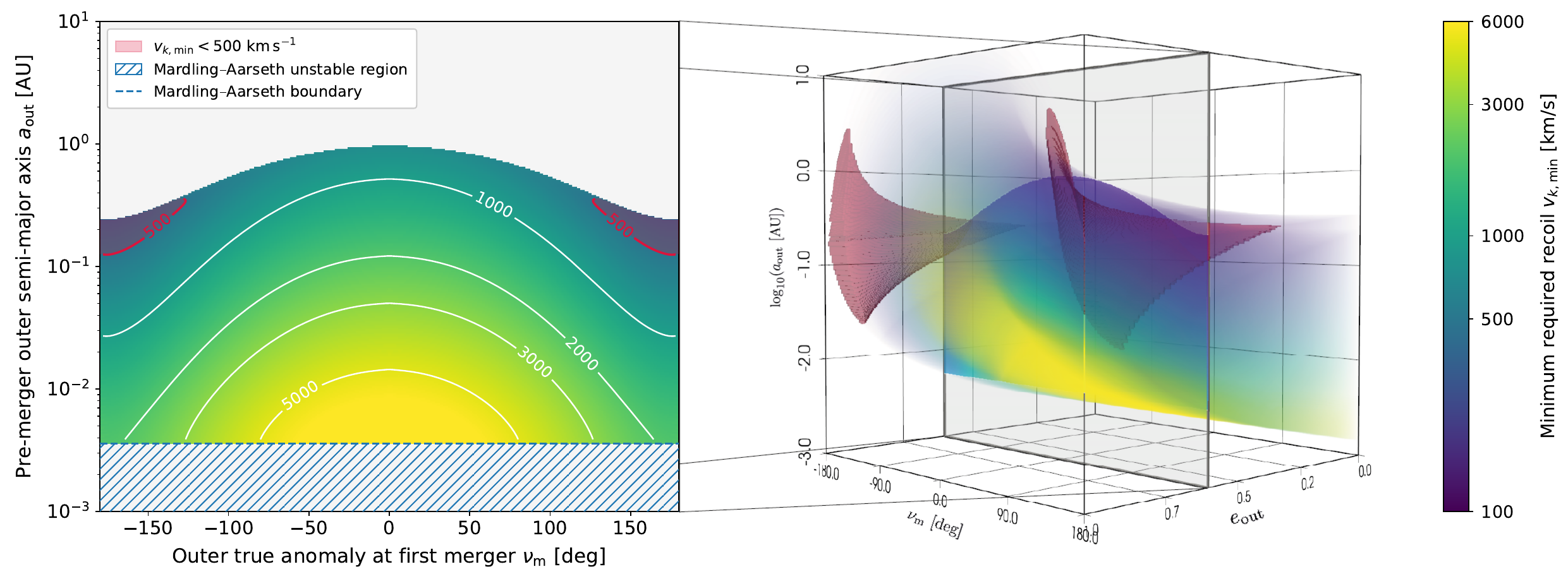}
    \caption{The image on the right shows a 3D volumetric rendering of a minimum-recoil-velocity map $v_{\rm k,min} \left( \boldsymbol{\theta}_{\rm out}; \boldsymbol{\lambda} \right)$ for an example system $\boldsymbol{\lambda}$. The recoil-compatible progenitor outer-orbit region $\mathcal A$, in which $v_{\rm k,min}
    \left(
        \boldsymbol{\theta}_{\rm out};
        \boldsymbol{\lambda}
    \right)
    \leq
    500~{\rm km\,s^{-1}}$, is colored in red. An $e_{\rm out}=const.$ slice is depicted on the left, with iso-contours for certain minimum recoil velocities. The minimum-recoil-velocity map is bounded from below by the Mardling-Aarseth criterion (i.e.~by the boundary of the physically admissible progenitor region $\mathcal D_{\rm phys}$), and from above by the boundary of the kinematically accessible progenitor region $\mathcal D_{\rm kin}$.}
    \label{fig:kick_map_concept}
\end{figure*}

\subsection{Minimum-recoil-velocity maps}
\label{sec:min_recoil_maps}

We next determine whether each target post-first-merger orbit can be produced from a given progenitor configuration $\boldsymbol{\theta}_{\rm out}
    =
    \left(
        a_{\rm out},
        e_{\rm out},
        \nu_{\rm m}
    \right)
    \in \mathcal D_{\rm phys}$.
The corresponding separation $R=r_{\rm out}(\boldsymbol{\theta}_{\rm out})$, radial velocity $v_{r,{\rm out}}(\boldsymbol{\theta}_{\rm out})$, and transverse velocity $v_{\phi,{\rm out}}(\boldsymbol{\theta}_{\rm out})$ at the time of the first merger are given by Eqs.~\eqref{eq:outer_separation}--\eqref{eq:outer_transverse_velocity}.

Because the first merger is treated as instantaneous, the separation is unchanged across the merger. A target post-first-merger orbit $(a_{\rm post},e_{\rm post})$ can therefore pass through the instantaneous separation $R$ only if $R$ lies between the post-first-merger pericenter and apocenter distances, which results in the condition
\begin{equation}
    a_{\rm post}(1-e_{\rm post})
    \leq
    R(\boldsymbol{\theta}_{\rm out})
    \leq
    a_{\rm post}(1+e_{\rm post}).
    \label{eq:radial_accessibility} 
\end{equation}
If this condition is not satisfied, the target orbit cannot be reached by any instantaneous velocity recoil at that progenitor point. This defines the set of accessible target orbits
\begin{equation}
    \begin{aligned}
        \mathcal T&
        \left(
            \boldsymbol{\theta}_{\rm out};
            \boldsymbol{\lambda}
        \right)
        =
        \Bigl\{
            (a_{\rm post},e_{\rm post})
            \in\mathcal T_{\rm obs}(\boldsymbol{\lambda})
            \,\Bigm| \\
            &a_{\rm post}(1-e_{\rm post})
            \leq
            R(\boldsymbol{\theta}_{\rm out})
            \leq
            a_{\rm post}(1+e_{\rm post})
        \Bigr\}.
    \end{aligned}
    \label{eq:accessible_target_orbits}
\end{equation}
We therefore define the domain of progenitor configurations admitting at least one kinematically accessible target orbit as
\begin{equation}
    \mathcal D_{\rm kin}
    \left(
        \boldsymbol{\lambda}
    \right)
    =
    \left\{
        \boldsymbol{\theta}_{\rm out}
        \in\mathcal D_{\rm phys}
        \,\middle|\,
        \mathcal T
        \left(
            \boldsymbol{\theta}_{\rm out};
            \boldsymbol{\lambda}
        \right)
        \neq\varnothing
    \right\}.
    \label{eq:kinematically_accessible_domain}
\end{equation}
For a kinematically accessible target orbit $(a_{\rm post},e_{\rm post})$, the required speed right after the first merger follows from the vis-viva relation,
\begin{equation}
    v_{\rm post}^2
    =
    GM
    \left(
        \frac{2}{R}
        -
        \frac{1}{a_{\rm post}}
    \right),
    \label{eq:post_vis_viva}
\end{equation}
while the magnitude of the specific orbital angular momentum is
\begin{equation}
    h_{\rm post}
    =
    \sqrt{
        GMa_{\rm post}
        \left(1-e_{\rm post}^2\right)
    }.
    \label{eq:post_h}
\end{equation}
The transverse speed is consequently
\begin{equation}
    v_{\phi,{\rm post}}
    =
    \frac{h_{\rm post}}{R},
    \label{eq:post_vphi}
\end{equation}
and the magnitude of the radial component is
\begin{equation}
    \left|v_{r,{\rm post}}\right|
    =
    \sqrt{
        v_{\rm post}^2-v_{\phi,{\rm post}}^2
    }.
    \label{eq:post_vr}
\end{equation}
The two signs of the radial component correspond to the outgoing and ingoing branches of the same Keplerian orbit. For fixed $(a_{\rm post},e_{\rm post})$, the smallest recoil is obtained when the post-merger transverse velocity remains in the progenitor orbital plane and is aligned with the pre-merger transverse motion. Any additional normal component increases the velocity difference without changing the required orbital energy or angular momentum. 

The minimum recoil velocity required to reach a particular target orbit $(a_{\rm post},e_{\rm post})$ from an initial progenitor outer-orbit configuration $\boldsymbol{\theta}_{\rm out}$ is therefore
\begin{equation}
\begin{aligned}
    &v_{\rm k}^2 \left(
        \boldsymbol{\theta}_{\rm out};
        a_{\rm post},e_{\rm post}
    \right)
    = \\
    &\min_{s=\pm1}
    \left[
        \left(
            s\left|v_{r,{\rm post}}\right|
            -
            v_{r,{\rm out}}
        \right)^2
        +
        \left(
            v_{\phi,{\rm post}}
            -
            v_{\phi,{\rm out}}
        \right)^2
    \right],
    \label{eq:target_minimum_recoil}
\end{aligned}
\end{equation}
where $s=+1$ and $s=-1$ correspond to the outgoing and ingoing branches, respectively.

In order to obtain the minimum recoil velocity $v_{\rm k,min}\left(\boldsymbol{\theta}_{\rm out};\boldsymbol{\lambda}\right)$ for a given system $\boldsymbol{\lambda}$ and a given progenitor configuration $\boldsymbol{\theta}_{\rm out} \in\mathcal D_{\rm kin}$, we minimize these recoil velocities over the complete accessible target orbit set,
\begin{equation}
    v_{\rm k,min}
    \left(
        \boldsymbol{\theta}_{\rm out};
        \boldsymbol{\lambda}
    \right)
    =
    \min_{
        (a_{\rm post},e_{\rm post})
        \in
        \mathcal T
        (\boldsymbol{\theta}_{\rm out};
         \boldsymbol{\lambda})
    }
    v_{\rm k}
    \left(
        \boldsymbol{\theta}_{\rm out};
        a_{\rm post},e_{\rm post}
    \right).
    \label{eq:min_recoil_target_set}
\end{equation}
For a fixed system $\boldsymbol{\lambda}$, this equation defines a minimum-recoil-velocity map over the kinematically admissible progenitor domain $\mathcal D_{\rm kin}$. 

We assume that the population of inner-binary black holes only produces recoil velocities below a specified recoil ceiling $v_{\rm k,max}$ and we therefore require
\begin{equation}
    v_{\rm k,min}
    \left(
        \boldsymbol{\theta}_{\rm out};
        \boldsymbol{\lambda}
    \right)
    \leq
    v_{\rm k,max}.
    \label{eq:recoil_cut}
\end{equation}
For each system $\boldsymbol{\lambda}$, this defines the recoil-compatible progenitor outer-orbit region
\begin{equation}
    \mathcal A
    \left(
        \boldsymbol{\lambda}; v_{\rm k,max}
    \right)
    =
    \left\{
        \boldsymbol{\theta}_{\rm out}
        \in\mathcal D_{\rm kin}
        \,\middle|\,
        v_{\rm k,min}
        \left(
            \boldsymbol{\theta}_{\rm out};
            \boldsymbol{\lambda}
        \right)
        \leq
        v_{\rm k,max}
    \right\}.
    \label{eq:allowed_progenitor_region}
\end{equation}
A representative minimum-recoil-velocity map for an example system with $v_{\rm k,max}=500 \text{ km s}^{-1}$ and a highlighted recoil-compatible progenitor region $\mathcal A$ is depicted in Fig.~\ref{fig:kick_map_concept}.

\subsection{Phase-space measure and constraint strength}
\label{sec:constraint_measures}

The surviving recoil-compatible fraction of progenitor parameter space depends on the measure assigned to that space. We express the outer semi-major axis in the dimensionless form
\begin{equation}
    x_{\rm out}
    =
    \frac{a_{\rm out}}{GM_{\rm pre}/c^2}.
    \label{eq:xout}
\end{equation} 
We adopt a parameter-space measure uniform in outer eccentricity, logarithmic dimensionless semi-major axis, and mean anomaly at merger $\mathcal M_{\rm m}$,
\begin{equation}
    {\rm d}\mu
    =
    {\rm d}e_{\rm out}\,
    {\rm d}\log_{10}x_{\rm out}\,
    {\rm d}\mathcal M_{\rm m}.
    \label{eq:phase_space_measure}
\end{equation}
A uniform mean-anomaly measure corresponds to uniform weighting in orbital time and is therefore a natural choice if the time of the first merger is not correlated with the outer orbital phase. In contrast, a uniform true-anomaly measure would assign equal weight to equal angular intervals despite the binary spending much less time near pericenter, thereby overrepresenting the pericenter region, particularly for highly eccentric outer orbits. We therefore evaluate the constraint strength using mean anomaly $\mathcal M_{\rm m}$, while constraints on the true anomaly $\nu_{\rm m}$ can be obtained analogously. The corresponding weighting is obtained from
\begin{equation}
    \frac{{\rm d}\mathcal M_{\rm m}}{{\rm d}\nu_{\rm m}}
    =
    \frac{
        \left(1-e_{\rm out}^2\right)^{3/2}
    }{
        \left(
            1+e_{\rm out}\cos\nu_{\rm m}
        \right)^2
    }.
    \label{eq:mean_true_anomaly_jacobian}
\end{equation}

Let $\mathcal R$ denote a general recoil-compatible progenitor region. At this stage the only such region available is $\mathcal A$ of Eq.~\eqref{eq:allowed_progenitor_region}. In Sec.~\ref{sec:peters_validity} we apply the definitions below to its secularly reliable subset $\mathcal A_{\rm sec}$. For the region $\mathcal R$, we define the surviving volume fraction
\begin{align}
    \mathcal{V}(\mathcal{R})
    &=
    \frac{
        \mu(\mathcal R)
    }{
        \mu(\mathcal D_{\rm phys})
    },
    \label{eq:fphys}
\end{align}
which quantifies the fraction of the physically admissible phase-space volume that remains compatible with the observed merger pair under the imposed recoil ceiling.

We further define the logarithmic constraint strength
\begin{equation}
    S(\mathcal{R})
    =
    -\log_{10}\mathcal{V}(\mathcal{R}).
    \label{eq:Svol}
\end{equation}
Thus, for example $S=1,\ 2,\ 3$
correspond to $10\%$, $1\%$, and $0.1\%$ of the physically admissible volume remaining allowed, respectively.

The total surviving volume does not by itself indicate which progenitor parameters are constrained. We therefore supplement $S$ with one-dimensional projection diagnostics. To avoid spurious isolated grid cells or numerical artifacts from contributing to these projections, we use a filtered projection of the allowed region. For a sampled value $X_j$ of a coordinate
\begin{equation}
    X
    \in
    \left\{
        e_{\rm out},
        \log_{10}x_{\rm out},
        \mathcal M_{\rm m}
    \right\},
\end{equation}
let $N_{\mathcal R}(X_j)$ denote the number of allowed cells in the transverse slice at $X_j$, and let $N_{\mathcal D}(X_j)$ denote the number of physically admissible cells in the same slice. We define the projection
\begin{equation}
    \Pi_X(\mathcal R)
    =
    \left\{
    X_j \,\middle|\,
    N_{\mathcal R}(X_j)\ge N_{\min},
    \quad
    \frac{N_{\mathcal R}(X_j)}
         {N_{\mathcal D}(X_j)}
    \ge \mathcal{V}_{\perp,\min}
    \right\},
    \label{eq:filtered_projection}
\end{equation}
where we set
\begin{equation}
    N_{\min}=2,
    \qquad
    \mathcal{V}_{\perp,\min}=10^{-4}.
\end{equation}
The first condition removes isolated single-cell features, while the second requires a minimum fractional occupancy of the physically admissible transverse slice. No contiguity or minimum interval length is imposed.

With this, we define the support coverage
\begin{equation}
    C_X^{\rm supp}(\mathcal{R})
    =
    \frac{
        L_X\left[\Pi_X(\mathcal R)\right]
    }{
        L_X\left[\mathcal I_X\right]
    },
    \label{eq:support_coverage}
\end{equation}
where $L_X[\Pi_X(\mathcal R)]$ is the total length of the interval(s) in $\Pi_X(\mathcal R)$ and $L_X[\mathcal I_X]$ is the length of the full scanned interval $\mathcal I_X$. Since the progenitor grid is uniform in $e_{\rm out}$, $\log_{10}x_{\rm out}$ and $\mathcal M_{\rm m}$, the measure of a region is proportional to its number of cells, and $\mathcal V$ and $C_X^{\rm supp}$ are evaluated by cell counting.

In addition to the continuous quantities $\mathcal{V}(\mathcal{R})$, $S(\mathcal{R})$, and $C_X^{\rm supp}(\mathcal{R})$, which constitute the quantitative results, we assign an operational constraint class according to the criteria in Table~\ref{tab:constraint_classes}. These categories are only intended for comparison within the atlas rather than as calibrated measures of statistical significance.

\begin{table}
    \centering
    \caption{
        Operational classification of the progenitor constraints for an allowed recoil-compatible progenitor region $\mathcal{R}$. The criteria are applied from top to bottom.}
    \label{tab:constraint_classes}
    \begin{tabular}{ll}
        \hline\hline
        Constraint Class & Criterion \\
        \hline
        Peters-incompatible
        &
        $\mathcal{V}(\mathcal{A}) = 0$
        \\
        Very strong
        &
        \begin{tabular}[t]{@{}l@{}}
            $\mathcal{V}(\mathcal{R})<0.01$ \\
            $\mathrm{and}\quad C_x^{\rm supp}(\mathcal{R})<0.20$ \\
            $\mathrm{and}\quad
            \min(C_e^{\rm supp}(\mathcal{R}),C_{\mathcal M}^{\rm supp}(\mathcal{R}))<0.30$
        \end{tabular}
        \\
        Strong
        &
        \begin{tabular}[t]{@{}l@{}}
            $\mathcal{V}(\mathcal{R})<0.03$ \\
            $\mathrm{and}\quad C_x^{\rm supp}(\mathcal{R})<0.30$ \\
            $\mathrm{and}\quad
            \min(C_e^{\rm supp}(\mathcal{R}),C_{\mathcal M}^{\rm supp}(\mathcal{R}))<0.50$
        \end{tabular}
        \\
        Moderate
        &
        \begin{tabular}[t]{@{}l@{}}
            $\mathcal{V}(\mathcal{R})<0.10$ \\
            $\mathrm{or}\quad C_x^{\rm supp}(\mathcal{R})<0.50$
        \end{tabular}
        \\
        Weak
        &
        otherwise
        \\
        \hline
    \end{tabular}
\end{table}

\subsection{Progenitor constraints for individual systems}
\label{sec:constraints_individual}

The preceding construction can be applied directly to an individual candidate merger pair. However, for individual systems, the quantities of primary interest are the locations and boundaries of the projected support intervals of the progenitor orbital parameters. Determining these ranges and boundaries can become challenging, especially when an allowed recoil-compatible region is very narrow or has a complicated geometry.

The general procedure for computing the orbital constraints can up to this point be summarized as follows. For a given system we determine the post-first-merger target orbits consistent with the prescribed inter-merger delay and detector-band eccentricity under Peters evolution. At each point on a regular grid in $(e_{\rm out},\log_{10}x_{\rm out},\mathcal M_{\rm m})$ within the physically admissible progenitor domain, we identify the kinematically accessible targets and minimize the required recoil over them to obtain $v_{\rm k,min}$. The phase grid is uniform in mean anomaly $\mathcal M_{\rm m}$, which is converted to true anomaly $\nu_{\rm m}$ at the corresponding eccentricity $e_{\rm out}$ when evaluating the orbital state. Imposing the recoil ceiling then defines the compatible progenitor region, from which we calculate the surviving volume fraction and projected support coverages introduced in Sec.~\ref{sec:constraint_measures}, as well as the locations and boundaries of the projected support intervals of the progenitor orbital parameters.

\begin{figure*}[htp!]
    \centering
    \vspace{-10pt}
    \includegraphics[width=0.95\textwidth]{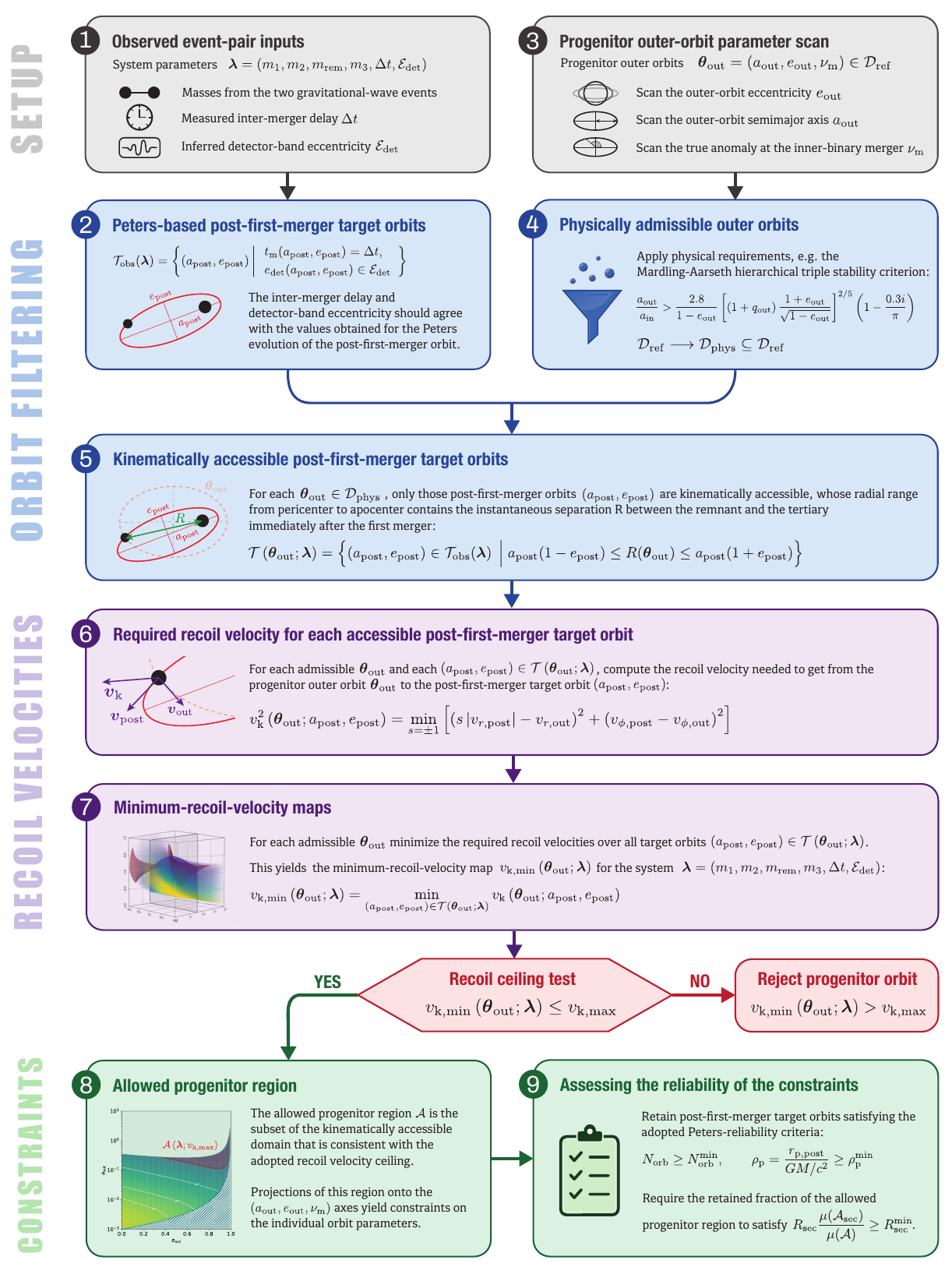}
    \caption{Schematic overview of the Peters-based progenitor-constraint
    construction. The inferred system properties
    $\boldsymbol{\lambda}$ define post-first-merger target orbits, while
    the progenitor parameters $\boldsymbol{\theta}_{\rm out}$ are scanned
    over the physically admissible domain. For each progenitor
    configuration, the recoil required to reach the kinematically
    accessible targets is minimized to obtain
    $v_{\rm k,min}(\boldsymbol{\theta}_{\rm out};
    \boldsymbol{\lambda})$. An upper recoil ceiling defines the allowed progenitor region and hence constraints on $a_{\rm out}$, $e_{\rm out}$, and $\nu_{\rm m}$. The schematic uses true anomaly $\nu_{\rm m}$, while numerical sampling and reported phase-support constraints use mean anomaly $\mathcal M_{\rm m}$.}
    \label{fig:method_chart}
\end{figure*}

The construction described so far includes all target orbits satisfying the inter-merger delay and detector-band-eccentricity conditions. Some of these target orbits correspond to only a small number of remaining orbital passages or to small pericenter separations, for which the orbit-averaged Peters approximation may no longer be reliable. In Sec.~\ref{sec:peters_validity}, we therefore introduce additional applicability criteria and quantify the extent to which the inferred recoil-compatible region is supported by targets satisfying them. The obtained constraints are only considered reliable for systems classified as Peters-reliable under these criteria. A concise flow-chart summary of the full procedure for computing the orbital constraints is illustrated in Fig.~\ref{fig:method_chart}.

For individual systems, we determine projected interval boundaries through local numerical refinement. We first compute the 3D minimum-recoil-velocity map at increasing resolutions (see Sec.~\ref{sec:convergence_tolerances} for the adopted resolutions and tolerances). Once the bulk convergence checks pass, including those on allowed volume fractions, support coverages, and $R_{\rm sec}$, we refine uncertain projected boundaries locally. At fixed orbital phase, for example, we scan eccentricity at successive semi-major axes, locate compatible regions, and refine their lower and upper eccentricity boundaries by bisection. We follow these boundaries across neighboring semi-major axes, using previous positions to guide additional sampling where the region becomes too narrow for the initial scan. Integrating the eccentricity widths over logarithmic semi-major axis and normalizing by the physically admissible area gives the support fraction at that phase. The procedure is repeated for the other parameters.

Apparent gaps between projected intervals are checked at several locations within each gap bin, repeating the edge-tracking calculation with finer sampling. Two successive calculations must agree on the support classification, unless all probes qualify as unambiguously empty after one round. If all probes consistently satisfy Eq.~\eqref{eq:filtered_projection}, the bin is added to the supported interval; if all remain below the threshold, the gap is retained. Mixed bins undergo boundary refinement, while inconclusive cases remain unresolved.

For endpoint refinement, we select parameter ranges around candidate boundaries on either comparison grid and bisect them, targeting bracket widths of half the adopted endpoint convergence tolerance. We report the bracket midpoint as the endpoint estimate and retain the bracket as its numerical uncertainty. For the semi-major axis, bisection and midpoint estimation are performed in $\log_{10}(a_{\rm out})$.

The method is implemented in our code \textsc{triple-check} \citep{triple-check}, which accepts fixed system parameters or posterior samples for two candidate events. Sample pairs are weighted using a truncated Gaussian matching kernel comparing the detector-frame remnant mass of the first event with the assigned component mass of the second, optionally also comparing logarithmic luminosity distance. We divide these weights by the second event’s marginal prior density in the matched coordinates to avoid counting the shared-coordinate prior twice. Pairs are drawn according to these weights, and the individual-system analysis is repeated for each distinct pair. Results are combined according to drawing frequency, yielding distributions of quantities such as the allowed volume fraction and projected support widths. For each progenitor coordinate, we also calculate the weighted fraction of sampled systems whose supported intervals contain that value. This quantifies how consistently a parameter value remains supported under measurement uncertainties. Further implementation details are provided in the code documentation.

\subsection{Constraint-atlas construction}
\label{sec:constraint_atlas}

To investigate systematically which combinations of component masses, inter-merger delays, and detector-band eccentricities yield informative constraints on the progenitor outer orbit, we construct an atlas over the intrinsic properties of the remnant-tertiary binary. We formulate the atlas in dimensionless variables and require every ingredient of the construction to be invariant under a common rescaling of all masses, lengths, and times. The atlas is then scale-free: one computation serves any reference frequency $f_{\rm ref}$, and fixing $f_{\rm ref}$ assigns each entry a definite total mass and inter-merger delay.

We introduce the dimensionless quantities
\begin{equation}
    \tau
    =
    \frac{c^3\Delta t}{GM},
    \qquad
    \widehat f_{\rm ref}
    =
    \frac{GMf_{\rm ref}}{c^3}.
    \label{eq:dimensionless_atlas_variables}
\end{equation}
At fixed $f_{\rm ref}$, $\widehat f_{\rm ref}$ is equivalent to specifying the total mass $M$ of the second-merger binary, while $\tau$ expresses the inter-merger delay in units of its gravitational timescale. These dimensionless quantities are unchanged by the conversion between source-frame and detector-frame quantities. Under this conversion, masses and time intervals acquire a factor of $1+z$, while frequencies acquire its inverse, so the redshift factors cancel in both $\tau$ and $\widehat f_{\rm ref}$. The corresponding source-frame and detector-frame descriptions therefore give the same dimensionless Keplerian and Peters evolution. Since gravitational-wave observations provide detector-frame masses together with the observed inter-merger delay and the observed reference frequency, the construction can be carried out consistently in the detector frame throughout. In particular, the merger time obtained from Eq.~(\ref{eq:peters_merger_time}) and the peak frequency of Eq.~(\ref{eq:wen_peak_frequency}) are themselves redshifted when evaluated with detector-frame masses, and can therefore be compared directly with the measured values. The dimensionless quantities $\widehat f_{\rm ref}$, $\tau$ and $x_{\rm out}$, as well as the recoil ceiling $v_{\rm k,max}$, are invariant under this rescaling, so that the constraint atlas does not depend on the redshift of the source. Consequently, the orbital and recoil construction itself can be expressed entirely in detector-frame quantities without requiring a source redshift. A redshift is required only when the inferred dimensionless separations are converted into source-frame physical separations, for which 
\begin{equation}
    a_{\rm out}^{\rm src}= \frac{a_{\rm out}^{\rm det}}{1+z}.
\end{equation}

Rather than retaining all component masses appearing in Eq.~\eqref{eq:general_system_parameters}, we characterize the post-first-merger binary by its total mass $M$ and symmetric mass ratio
\begin{equation}
    \eta
    =
    \frac{m_{\rm rem}m_3}{M^2}.
    \label{eq:eta_atlas}
\end{equation}
For specified $M$ and $\eta$, the two component masses are
\begin{equation}
    m_{\pm}
    =
    \frac{M}{2}
    \left(
        1
        \pm
        \sqrt{1-4\eta}
    \right).
    \label{eq:atlas_component_masses}
\end{equation}
Because the recoil is imparted specifically to the first-merger remnant, interchanging the remnant and tertiary masses does not in general yield an equivalent progenitor system. We therefore evaluate both mass-ordering branches,
\begin{align}
    \mathcal B_{>}
    &:\
    (m_3,m_{\rm rem})
    =
    (m_+,m_-),
    \label{eq:branch_greater}
    \\
    \mathcal B_{<}
    &:\
    (m_3,m_{\rm rem})
    =
    (m_-,m_+).
    \label{eq:branch_less}
\end{align}

For the baseline atlas, we adopt a fixed fractional mass loss of
\begin{equation}
    \epsilon_{\rm rad}=0.05.
\end{equation}
Numerical-relativity calculations and phenomenological fits typically predict that a few percent of the total binary mass is radiated in gravitational waves for comparable-mass black-hole mergers \citep{Barausse2012}. We note that $\epsilon_{\rm rad}$ and the recoil velocity are both determined by the inner-binary mass ratio and spin configuration and are therefore generally correlated. Our treatment of a fixed $\epsilon_{\rm rad}$ together with a freely varying $\boldsymbol{v}_k$ neglects this correlation and should be understood as a simplifying approximation. Using this fixed fractional mass loss, the other relevant quantities can be computed using
\begin{equation}
    m_{12} =\frac{m_{\rm rem}}{1- \epsilon_{\rm rad}}, \quad q_{\rm out} = \frac{m_3}{m_{12}} = (1- \epsilon_{\rm rad}) \frac{m_3}{m_{\rm rem}}.
\end{equation}
We repeat selected calculations with $\epsilon_{\rm rad}=0.03$ and
$0.08$ to assess the sensitivity to this assumption. 

A point in the intrinsic constraint atlas is specified by
\begin{equation}
    \boldsymbol{\Theta}
    =
    \left(
        \widehat f_{\rm ref},
        \eta,
        \tau,
        e_{\rm ref,c},
        \mathcal B
    \right),
    \label{eq:intrinsic_atlas_vector}
\end{equation}
where $\mathcal B\in\{\mathcal B_{>},\mathcal B_{<}\}$ denotes the mass-ordering branch and where $e_{\rm ref,c}$ is the central eccentricity at the reference frequency $f_{\rm ref}$ for which we choose a fiducial half-width
\begin{equation}
    \delta e=0.01.
\end{equation} 
Our initial broad atlas scan covers
\begin{align}
    10^{-3}
    &\leq
    \widehat f_{\rm ref}
    \leq
    1.5\times10^{-2},
    \label{eq:broad_scan_f}
    \\
    0.05
    &\leq
    \eta
    \leq
    0.25,
    \\
    10^7
    &\leq
    \tau
    \leq
    10^{12},
    \\
    0
    &\leq
    e_{\rm ref,c}
    \leq
    0.3,
    \label{eq:broad_scan_e}
\end{align}
At $f_{\rm ref}=10\,\mathrm{Hz}$, the frequency range corresponds to total detector-frame masses $M\simeq20$--$305\,M_\odot$. The initial frequency and delay coordinates are sampled logarithmically, while $\eta$ is sampled linearly. For $e_{\rm ref,c}$, the initial grid is
\begin{equation}
    e_{\rm ref,c}
    =
    \{0,0.01,0.03,0.05,0.10,0.20,0.30\},
\end{equation}
providing denser sampling at low eccentricity. Both mass-ordering branches are evaluated at every unequal-mass point. For $\eta=0.25$ the two branches coincide and only one independent system is evaluated.

For all atlas systems we choose the rectangular reference domain
\begin{equation}
    \mathcal D_{\rm ref}
    =
    \left\{
        \left(
            e_{\rm out},x_{\rm out},\nu_{\rm m}
        \right)
        \,\middle|\,
        \begin{array}{l}
        0\leq e_{\rm out} \leq 0.999,\\
        10^2\leq x_{\rm out}\leq10^8,\\
        -\pi\leq\nu_{\rm m}<\pi
        \end{array}
    \right\}
    \label{eq:Dref}
\end{equation}
of the associated minimum-recoil-velocity maps. This range serves as a standardized reference domain rather than an astrophysical prior. All systems whose dynamically allowed region reaches $x_{\rm out} = 10^2$ or $x_{\rm out} = 10^8$ are recalculated with an extended separation range up to $x_{\rm out} = 10^1$ or $x_{\rm out} = 10^9$, respectively. Representative systems whose allowed region reaches $e_{\rm out} = 0.999$ are additionally repeated with $e_{\rm out} \leq 0.9999$. These extensions are used to assess the sensitivity of the inferred constraints to the adopted parameter-space boundaries.

The physically admissible domain $\mathcal D_{\rm phys}$ is defined by imposing the hierarchical-stability condition introduced in Sec.~\ref{sec:physical_domain}. The Mardling-Aarseth condition \eqref{eq:ma_stability_criterion} is evaluated at a fiducial dimensionless reference separation
\begin{equation}
    \alpha_{\rm in} \equiv \frac{a_{\rm in}}{Gm_{12}/c^2}=700.
    \label{eq:alpha_in}
\end{equation}
We also repeat the calculations for selected systems with $\alpha_{\rm in}=350$ and $\alpha_{\rm in}=1400$ to assess the sensitivity to this assumption.

Because the outer separation is measured in units of the pre-merger total mass, $x_{\rm out}=a_{\rm out}/(GM_{\rm pre}/c^2)$ with $M_{\rm pre}=m_{12}+m_3$, we have $a_{\rm in}/(GM_{\rm pre}/c^2)=\alpha_{\rm in}/(1+q_{\rm out})$. After substituting $\alpha_{\rm in} = 700$ into Eq.~\eqref{eq:aoutmin_MA}, the minimum stable outer semi-major axis can be written as
\begin{equation}
    x_{\rm out,min}^{\rm MA}
    =
    1372
    \frac{\left(1+e_{\rm out}\right)^{2/5}}
         {\left(1+q_{\rm out}\right)^{3/5} \left(1-e_{\rm out}\right)^{6/5}}
    \label{eq:xoutmin_MA}
\end{equation}
in dimensionless variables.

To convert the minimum-recoil maps into progenitor constraints, we further adopt the recoil ceiling
\begin{equation}
    v_{\rm k,max}=500~{\rm km\,s^{-1}}.
\end{equation}
This value is motivated by numerical-relativity simulations of nonprecessing black-hole binaries. For spins aligned or antialigned with the orbital angular momentum, Ref.~\cite{Healy2014} finds a maximum recoil of $526\pm23~{\rm km\,s^{-1}}$, while nonspinning unequal-mass binaries attain considerably smaller maximum recoils of approximately $175~{\rm km\,s^{-1}}$ \citep{Gonzalez2007}. The adopted value should not, however, be interpreted as a universal maximum: generic precessing configurations with large in-plane spins can produce recoils approaching $5000~{\rm km\,s^{-1}}$ \citep{Lousto2011}. We therefore regard $v_{\rm k,max}=500~{\rm km\,s^{-1}}$ as a fiducial, population-dependent ceiling rather than a fundamental bound.

With all these parameters and assumptions specified, we construct the
constraint atlas by applying the procedure from Sec.~\ref{sec:constraints_individual} and Fig.~\ref{fig:method_chart} to each individual system in the atlas. For atlas systems we do not employ the local refinement strategies to get accurate endpoint values for the projected support intervals, since only the volume fractions and support coverages are relevant for determining the constraint strength. 

\section{Applicability of the Peters approximation and post-Newtonian validation}
\label{sec:peters_validity}

The constraint construction introduced in Sec.~\ref{sec:peters_constraints} uses the orbit-averaged Peters equations to relate the orbital elements of the post-first-merger binary to its merger time and detector-band eccentricity. This treatment assumes that the orbital elements evolve sufficiently slowly compared with the orbital timescale. The approximation can become inaccurate for highly eccentric or compact binaries, for which a substantial fraction of the orbital energy and angular momentum may be radiated during few individual pericenter passages \citep{Peters1964,Zwick2020,Fumagalli2025}.

We therefore do not regard every solution of the Peters equations as equally reliable. Instead, we associate each post-first-merger target orbit with diagnostics that quantify whether its subsequent evolution is expected to remain sufficiently adiabatic for the orbit-averaged description to provide a useful approximation.

\subsection{Applicability diagnostics}
\label{sec:peters_diagnostics}

For a target orbit $(a_{\rm post},e_{\rm post})$, we first define its initial Keplerian orbital period,
\begin{equation}
    P_{\rm post}
    =
    2\pi
    \sqrt{
        \frac{a_{\rm post}^3}{GM}
    },
    \label{eq:Ppost}
\end{equation}
and introduce the characteristic number of remaining orbital periods,
\begin{equation}
    N_{\rm orb}
    =
    \frac{\Delta t}{P_{\rm post}}.
    \label{eq:Norb}
\end{equation}
This quantity is not intended as an exact count of the accumulated orbital cycles. Because the orbital period decreases, $N_{\rm orb}$ provides a conservative estimate of the number of orbital periods available for the orbit-averaged evolution.

A complementary diagnostic is provided by the initial post-merger pericenter distance,
\begin{equation}
    r_{\rm p,post}
    =
    a_{\rm post}
    \left(
        1-e_{\rm post}
    \right),
    \label{eq:rp_post}
\end{equation}
which we express in units of the gravitational radius as
\begin{equation}
    \rho_{\rm p}
    =
    \frac{r_{\rm p,post}}{GM/c^2}.
    \label{eq:rho_p}
\end{equation}
Small values of $\rho_{\rm p}$ indicate that the binary undergoes strong-field pericenter passages, for which the weak-field and orbit-averaged assumptions underlying the Peters evolution become increasingly questionable. For our analysis, we classify the Peters equations as applicable to a post-first-merger target orbit if
\begin{equation}
    N_{\rm orb}
    \geq
    10,
    \qquad
    \rho_{\rm p}
    \geq
    20.
    \label{eq:peters_validity_cuts}
\end{equation}

\begin{figure*}
    \centering
    \includegraphics[width=1\textwidth]{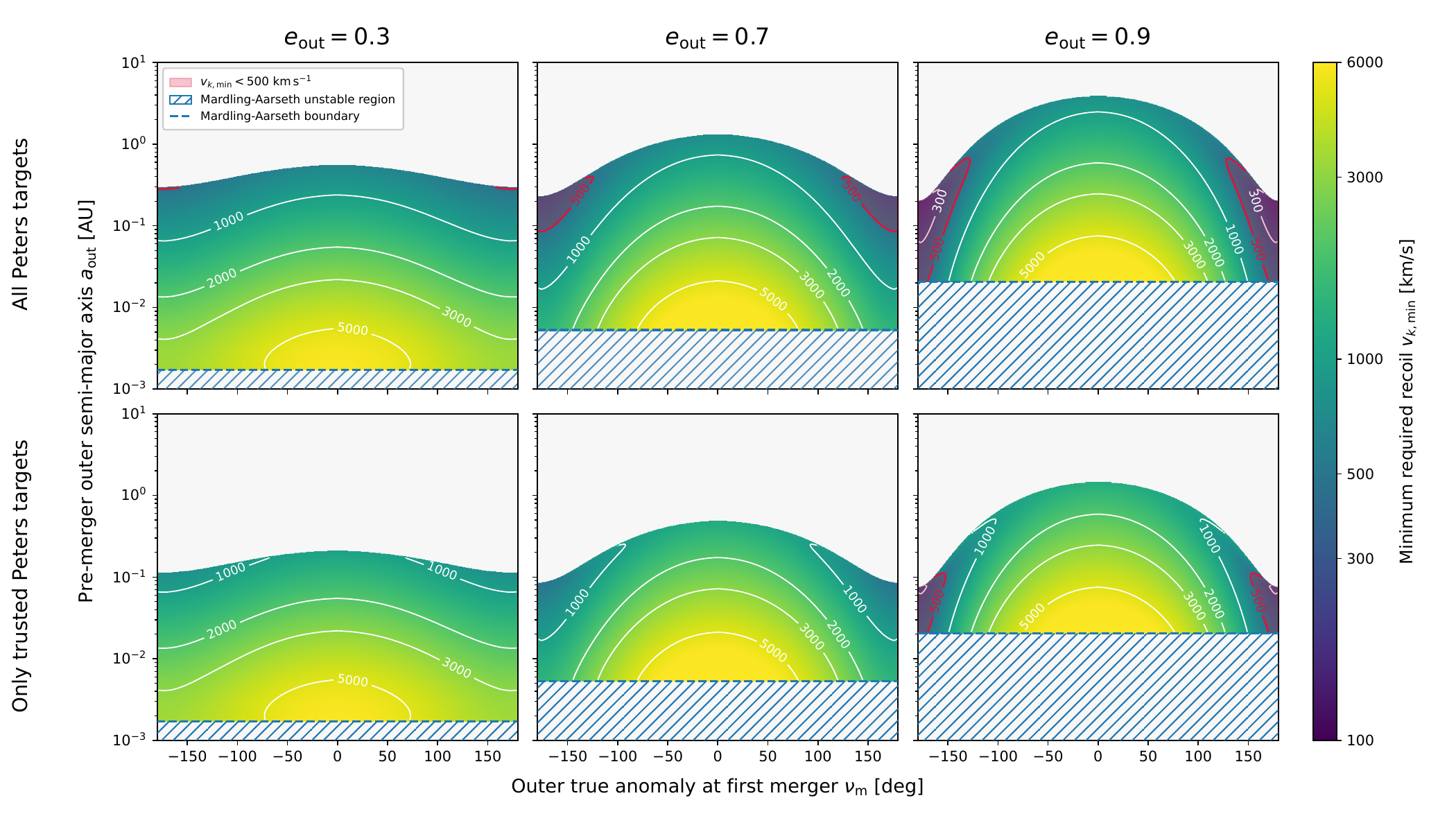}
    \vspace{-20pt}
    \caption{Comparison of minimum-recoil-velocity maps for several fixed values of $e_{\rm out}$ for a representative example system. The top row shows $v_{k,\min}$ obtained by minimizing over all Peters target orbits within $\mathcal{D}_{\rm kin}$, while the bottom row includes only targets satisfying the adopted Peters-applicability criteria from Eq.~\eqref{eq:peters_validity_cuts}, restricting the domain to $\mathcal{D}_{\rm sec}$. For this example, the minimum-recoil-velocity map is strongly influenced by nonsecular targets that fail the Peters-applicability criteria.}
    \label{fig:secular_comparison}
\end{figure*}

For each progenitor configuration $\boldsymbol{\theta}_{\rm out}$, we define the secularly reliable subset of the kinematically accessible targets as
\begin{equation}
\begin{aligned}
    \mathcal T_{\rm sec} &
    \left(
        \boldsymbol{\theta}_{\rm out};
        \boldsymbol{\lambda}
    \right)
    = \left\{
        (a_{\rm post},e_{\rm post})
        \in
        \mathcal T
        \left(
            \boldsymbol{\theta}_{\rm out};
            \boldsymbol{\lambda}
        \right)
        \,\middle|\,
        \begin{array}{l}
        N_{\rm orb}\geq 10,\\
        \rho_{\rm p}\geq 20
        \end{array}
    \right\}.
    \label{eq:Tsec}
\end{aligned}
\end{equation}
The corresponding progenitor configurations for which at least one such target exists form the domain
\begin{equation}
    \mathcal D_{\rm sec}
    \left(
        \boldsymbol{\lambda}
    \right)
    =
    \left\{
        \boldsymbol{\theta}_{\rm out}
        \in
        \mathcal D_{\rm phys}
        \,\middle|\,
        \mathcal T_{\rm sec}
        \left(
            \boldsymbol{\theta}_{\rm out};
            \boldsymbol{\lambda}
        \right)
        \neq
        \varnothing
    \right\}.
    \label{eq:Dsec}
\end{equation}
For these configurations, the minimum recoil velocity is recalculated by minimizing only over the secularly reliable targets,
\begin{equation}
    v_{\rm k,min}^{\rm sec}
    \left(
        \boldsymbol{\theta}_{\rm out};
        \boldsymbol{\lambda}
    \right)
    =
    \min_{
        (a_{\rm post},e_{\rm post})
        \in
        \mathcal T_{\rm sec}
        (
            \boldsymbol{\theta}_{\rm out};
            \boldsymbol{\lambda}
        )
    }
    v_{\rm k}
    \left(
        \boldsymbol{\theta}_{\rm out};
        a_{\rm post},
        e_{\rm post}
    \right).
    \label{eq:vkmin_sec}
\end{equation}
The secularly reliable recoil-compatible progenitor region is then
\begin{equation}
    \mathcal A_{\rm sec}
    \left(
        \boldsymbol{\lambda};
        v_{\rm k,max}
    \right)
    =
    \left\{
        \boldsymbol{\theta}_{\rm out}
        \in
        \mathcal D_{\rm sec}
        \,\middle|\,
        v_{\rm k,min}^{\rm sec}
        \left(
            \boldsymbol{\theta}_{\rm out};
            \boldsymbol{\lambda}
        \right)
        \leq
        v_{\rm k,max}
    \right\}.
    \label{eq:Asec}
\end{equation}

To assess whether the Peters-based progenitor constraints for a given system are themselves robust, we compare the unrestricted allowed region with its secularly reliable subset. We introduce the retained reliable fraction
\begin{equation}
    R_{\rm sec}
    =
    \frac{\mathcal{V}(\mathcal A_{\rm sec})}{\mathcal{V}(\mathcal A)}
    =
    \frac{
        \mu(\mathcal A_{\rm sec})
    }{
        \mu(\mathcal A)
    },
    \qquad \mathcal{V}>0.
    \label{eq:Rsec}
\end{equation}
A system is classified as \emph{Peters-reliable} if
\begin{equation}
    R_{\rm sec}
    \geq
    0.85.
    \label{eq:Rsec_cut}
\end{equation}
In this case, the reported constraint measures are evaluated from the secularly reliable region $\mathcal A_{\rm sec}$. Systems with $\mathcal{V}>0$ but $R_{\rm sec}<0.85$ are classified as \emph{Peters-unreliable}. If $\mathcal{A}=\varnothing$, the retained fraction $R_{\rm sec}$ is undefined and the system lies outside the Peters-reliable/Peters-unreliable classification. We describe such a system as \emph{Peters-incompatible}, meaning that no progenitor within the adopted domain reproduces the prescribed inter-merger delay and detector-band eccentricity under Peters evolution while satisfying the imposed recoil ceiling. This is a model-dependent exclusion rather than a definitive rejection of a hierarchical-triple origin. A non-orbit-averaged treatment can shift the family of post-merger target orbits and may therefore create or remove compatible solutions. A stronger exclusion would require a numerically converged and domain-extended inverse search repeated using post-Newtonian or more complete relativistic dynamics.

Fig.~\ref{fig:secular_comparison} illustrates the possible impact of imposing the aforementioned applicability criteria. For the depicted example system, the inferred allowed region of progenitor outer-orbit parameters is strongly affected by post-first-merger target orbits that fail these criteria. The Peters-based reconstruction is consequently not sufficiently reliable for this specific system.

\subsection{Post-Newtonian validation}
\label{sec:pn_validation}

The applicability cuts above provide inexpensive diagnostics that can be evaluated throughout the full constraint atlas, but they do not constitute a direct test of the Peters approximation. We therefore perform post-Newtonian (PN) integrations for representative configurations spanning the range of applicability diagnostics encountered in the atlas. In particular, we select systems lying well within the secularly reliable region as well as systems approaching the boundaries in $N_{\rm orb}$ and $\rho_{\rm p}$.

For each selected progenitor configuration, we use the same instantaneous mass-loss and recoil mapping as in Sec.~\ref{sec:first_merger_mapping} to construct the post-first-merger relative position and velocity, $\left(\boldsymbol r_{\rm post}, \boldsymbol v_{\rm post} \right)$. These quantities, together with the anticipated values $(a_{\rm post}, e_{\rm post})$ of the target orbit, are then used to construct initial data for a direct post-Newtonian evolution of the remnant--tertiary binary. The equations of motion include Newtonian gravity, conservative corrections at 1PN and 2PN order, and the leading-order radiation-reaction term at 2.5PN order, resulting in the Hamiltonian
\begin{equation}
    H = H_{\rm N} + H_{\rm 1PN} + H_{\rm 2PN} + H_{\rm 2.5PN}.
\end{equation}

For the numerical integration of the corresponding equations of motion we use the \textsc{pn-nbody} code \cite{pn-nbody} with an embedded 5th-order Runge-Kutta method to achieve a relative estimated error tolerance of $10^{-14}$ and we terminate the dynamical evolution at a separation of $2GM/c^2$. For more details on the numerical methods and the post-Newtonian formalism used in the code, see Ref.~\cite{Heinze2026}.

Detector-band eccentricities are evaluated at a reference frequency of $10\,\mathrm{Hz}$, using Wen's peak-frequency prescription with the instantaneous PN radial elements. The frequency crossing is resolved through a finely sampled restart. We quantify discrepancies by
\begin{equation}
    \delta_t
    =
    \frac{
        t_{\rm m}^{\rm PN}
        -
        t_{\rm m}^{\rm Peters}
    }{
        t_{\rm m}^{\rm Peters}
    },
    \label{eq:delta_t_PN}
\end{equation}
and 
\begin{equation}
    \delta e_{\rm ref} = e_{r,\rm PN}(10\,\mathrm{Hz})-e_{\rm Peters}(10\,\mathrm{Hz}),
\end{equation}
where $e_r=(r_a-r_p)/(r_a+r_p)$ is obtained from the conservative radial turning points. 

For representative systems and selected points within their allowed recoil-compatible regions $\mathcal A_{\rm sec}$, we additionally evolve the corresponding minimum-recoil configurations and determine whether they remain consistent with the prescribed inter-merger delay $\Delta t$ and detector-band eccentricity $\mathcal E_{\rm ref}$ under PN evolution. This provides a direct test of the quantities entering the progenitor constraints rather than only of the underlying orbital-evolution model.

\section{Results}
\label{sec:results}

\subsection{Robustness of the atlas construction}
\label{sec:atlas_robustness}

Before constructing the full constraint atlas, we first generate a lower-resolution atlas covering the full intrinsic parameter range described in Eqs.~\eqref{eq:broad_scan_f}-\eqref{eq:broad_scan_e}. The base grid contains seven frequency, seven symmetric-mass-ratio, nine delay, and seven eccentricity values. Removing the redundant equal-mass branch leaves 5733 independent base systems. We use these systems to assess the numerical stability of the inferred constraints, their sensitivity to the reference domain and baseline modeling assumptions, and perform selected comparisons with post-Newtonian simulations.

\subsubsection{Numerical stability and tolerances}
\label{sec:convergence_tolerances}
We first perform convergence tests on representative allowed
progenitor regions selected from the preliminary atlas, in particular on exceptionally narrow regions consisting of relatively few cells. Agreement between two resolutions requires the retained fraction $R_{\rm sec}$ and each projected support coverage to change by at most $0.01$ in absolute value, and both the all-target and secular allowed volume fractions to change by at most $0.01$ dex. A zero/nonzero transition fails this criterion. Based on these tests, we adopt a strategy in which each system starts from a survey map with resolution $(N_e,N_x,N_M,N_{e_{\rm post}})=(75,270,270,3750)$ and is then advanced independently through an increasing resolution ladder until its numerical agreement criteria are satisfied or the prescribed resolution limit is reached. The full resolution sequence can be found in Table \ref{tab:resolution_sequence}. Ordinary systems require agreement between one consecutive pair of resolutions. Stronger confirmation is imposed for thin recoil-compatible regions, systems near the Peters-reliability threshold, and maps without a recoil-compatible region. A region is flagged as thin if the cell count is at most 100, or if the region occupies at most two projected bins in eccentricity or semi-major axis. Proximity to the reliability threshold is defined by $|R_{\rm sec}-0.85|\leq0.001$. If either flag occurs in the latest three maps, both adjacent comparisons among three consecutive resolutions must pass. An additional calculation at the finest resolution shifts the eccentricity and logarithmic semi-major axis nodes to interval midpoints and offsets the phase grid by half a cell, while retaining the reference domain and target sampling. This shifted calculation must agree with the finest uniform map under the same criteria. Using this approach, all of our base atlas systems satisfy our numerical stability criteria. For individual-system summaries, the default convergence tolerances for the projected support boundaries are 0.01 in $e_{\rm out}$, 0.01 dex in $\log_{10}a_{\rm out}$, and $1^\circ$ in orbital phase. 

\begin{table}
\centering
\caption{Resolutions used for the convergence tests.}
\label{tab:resolution_sequence}
\begin{tabular*}{0.85\linewidth}{@{\extracolsep{\fill}}ccccc@{}}
\hline
Level & $N_e$ & $N_x$ & $N_M$ & $N_{e_{\rm post}}$ \\
\hline
1 &  75 & 270 & 270 &  3750 \\
2 & 100 & 360 & 360 &  5000 \\
3 & 125 & 450 & 450 &  6250 \\
4 & 150 & 540 & 540 &  7500 \\
5 & 200 & 720 & 720 & 10000 \\
6 & 250 & 900 & 900 & 12500 \\
\hline
\end{tabular*}
\end{table}

\subsubsection{Sensitivity to the reference domain}

We additionally assess the sensitivity of the base atlas to the reference domain given in Eq.~\eqref{eq:Dref} by extending individual boundaries for Peters-reliable systems whose allowed regions reach those boundaries. Increasing $e_{\rm out,max}$ from $0.999$ to $0.9999$ affects 196 base-grid systems. None changes recoil compatibility or Peters-reliability status, and only one changes from weakly to moderately constrained. The median relative change in the secular allowed fraction is $+0.23\%$, and the maximum absolute relative change is approximately $0.34\%$. 27 regions still reach the extended boundary. No sampled allowed region touches $x_{\rm out,min}=100$, so no lower-boundary extension is needed. Increasing $x_{\rm out,max}$ from $10^8$ to $10^9$ affects 123 base-grid systems. Compatibility and Peters-reliability status remain unchanged, but 33 systems change from weakly to moderately constrained, while the median normalized secular allowed fraction decreases by $18.4\%$. Of these regions, 119 still reach the extended upper boundary. Because the fractions and support coverages are normalized within each run's domain, these class changes do not necessarily indicate stronger physical constraints. No baseline strong or very-strong base-grid constraint touches any of the tested boundaries. These results support the stability of the sampled constraints against the eccentricity extension, while demonstrating continued upper-radius dependence for weak and moderate constraints. We therefore retain $x_{\rm out,max}=10^8$ as our reference cutoff and interpret the reported fractions and classifications relative to this domain.

\begin{figure*}
    \centering
    \includegraphics[width=1\textwidth]{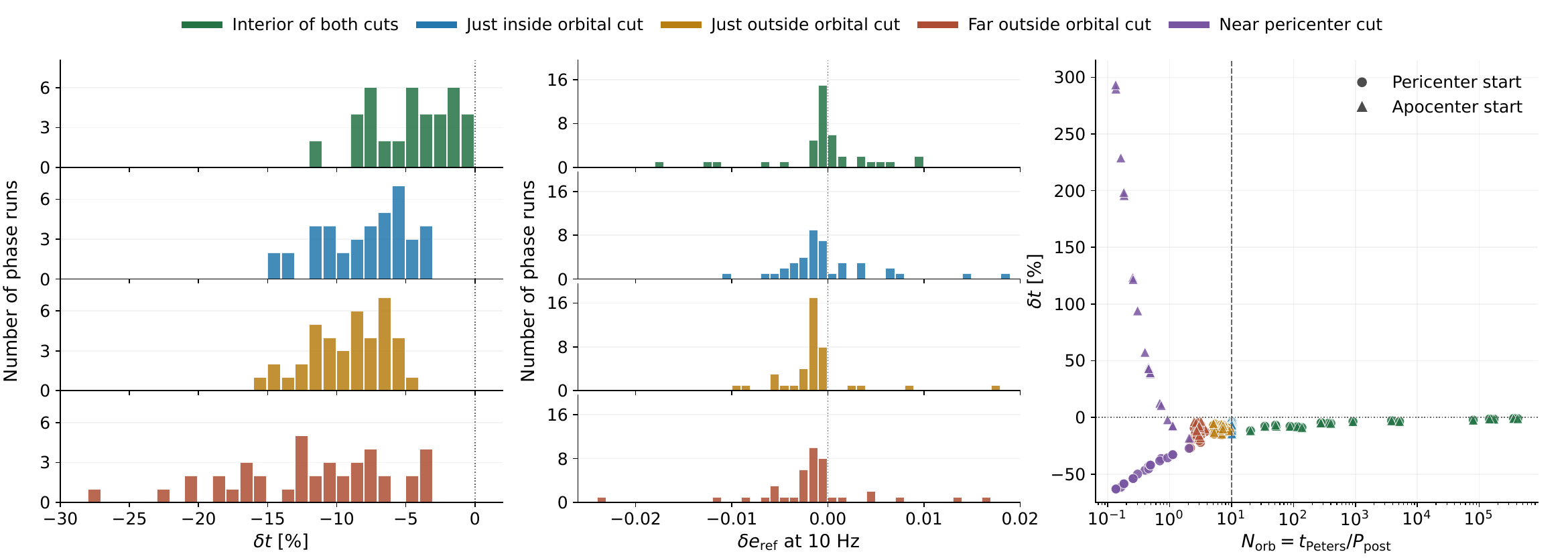}
    \caption{Comparison of Peters predictions with direct 2.5PN integrations for 100 atlas-selected binaries, each initialized at pericenter and apocenter. Left and middle: distributions of the merger-time discrepancies $\delta t$ and the eccentricity discrepancies $\delta e_{\rm ref}$ at $10\,\mathrm{Hz}$, for four different diagnostic categories. Right: merger-time discrepancy versus $N_{\rm orb}$ for all five categories. Circles and triangles indicate pericenter and apocenter starts, respectively; the dashed vertical line marks the threshold $N_{\rm orb}=10$.}
\label{fig:pn_validation}
\end{figure*}

\subsubsection{Dependence on the baseline modeling assumptions}

Finally, we assess how the inferred constraints change under variations of the baseline modeling assumptions. For this, we use a subset of the broad preliminary atlas, retaining all seven frequency nodes, four symmetric-mass-ratio nodes, seven merger delay nodes, and six eccentricity nodes. Including both mass-ordering branches at unequal masses and removing the redundant equal-mass branch gives 2058 independent systems. Varying the fractional mass radiated in the first merger from $\epsilon_{\rm rad}=0.05$ to $0.03$ or $0.08$ has only a weak effect: the number of Peters-reliable, recoil-compatible systems on the reduced sensitivity grid changes from 210 to 213 for $\epsilon_{\rm rad}=0.03$ and stays at 210 for $\epsilon_{\rm rad}=0.08$, while the typical allowed progenitor volumes of systems retained in both calculations change by only $+1.2\%$ and $-1.9\%$, respectively. Changing the dimensionless inner separation from $\alpha_{\rm in}=700$ to $350$ or $1400$ changes the number of Peters-reliable, recoil-compatible systems from 210 to 256 and 179, respectively. For systems retained in both calculations, the median allowed progenitor volume changes by approximately $+18\%$ and $-14\%$. The dominant modeling dependence arises from the adopted recoil ceiling, which is expected. Relative to $v_{\rm k,max}=500\,{\rm km\,s^{-1}}$, reducing the ceiling to $300\,{\rm km\,s^{-1}}$ decreases the number of Peters-reliable, recoil-compatible systems from 210 to 161. For systems retained in both scans, the allowed progenitor volume is typically reduced by a factor of approximately $3.3$. Increasing the ceiling to $1000\,{\rm km\,s^{-1}}$ increases the number of retained systems to 298 and expands the typical allowed volume of systems common to both scans by a factor of approximately $4.7$.

\subsubsection{Post-Newtonian validation}
\label{sec:pn_validation_results}

We assess the Peters-based predictions using direct integrations of 100 distinct atlas-selected binaries with the methods described in Sec.~\ref{sec:pn_validation}. We select 20 binaries in each of five diagnostic categories: the interior of both Peters-applicability cuts from Eq.~\eqref{eq:peters_validity_cuts}, just inside the orbital cut, just outside the orbital cut, far outside the orbital cut, and near the pericenter cut. All selected binaries in the last category also violate the orbital condition, with $N_{\rm orb}\simeq0.13$--$2.06$. This category therefore probes the combined breakdown of the two reliability conditions rather than isolating the pericenter criterion. Within each category, sampling is distributed across available logarithmic delay bins with $\tau\leq10^{11}$, giving physical Peters lifetimes between approximately $0.07$ and $252$ days. For our main comparisons, the initial PN radial semi-major axis and eccentricity are matched to the corresponding Peters orbital elements. Each binary is initialized separately at pericenter and apocenter, yielding 200 integrations in total.

In the first four categories, PN merger times are systematically shorter than the Peters predictions, with median absolute discrepancies of $4.71\%$, $7.21\%$, $8.57\%$, and $10.96\%$, respectively. Their median absolute eccentricity discrepancies are $8.3\times10^{-4}$, $2.1\times10^{-3}$, $1.5\times10^{-3}$, and $2.0\times10^{-3}$. The gradual increase in typical lifetime discrepancies towards smaller $N_{\rm orb}$ indicates that the orbital cut provides a practical applicability criterion rather than a sharp boundary between accurate and inaccurate predictions. The eccentricity discrepancies do not exhibit a similarly monotonic trend. The near-pericenter category exhibits substantially larger lifetime discrepancies, ranging from $-63.3\%$ to $+293.3\%$, with a median absolute discrepancy of $43.96\%$. Initial-phase sensitivity also increases substantially. This behavior illustrates the limitations of a phase-independent, orbit-averaged description when the predicted lifetime is comparable to or shorter than the initial orbital period. The results are summarized in Fig.~\ref{fig:pn_validation}. Across the full sample, 178 of 200 integrations satisfy $|\delta e_{\rm ref}|\leq0.01$. 50 additional tests for representative minimum-recoil configurations from allowed recoil-compatible regions agree with the previous tests, exhibiting median absolute merger time discrepancies of $5.53\%$ and median absolute eccentricity discrepancies of $6.1\times10^{-3}$.

Taken together, all these tests support the numerical stability and robustness of the atlas systems under the adopted criteria, while identifying sensitivity to the reference domain and recoil ceiling.

\subsection{Constraint atlas}
\label{sec:constraint_atlas_results}

\subsubsection{Peters-reliable region}
\label{sec:peters_reliable_region}

\begin{figure*}[t]
    \centering
    \includegraphics[width=\textwidth]{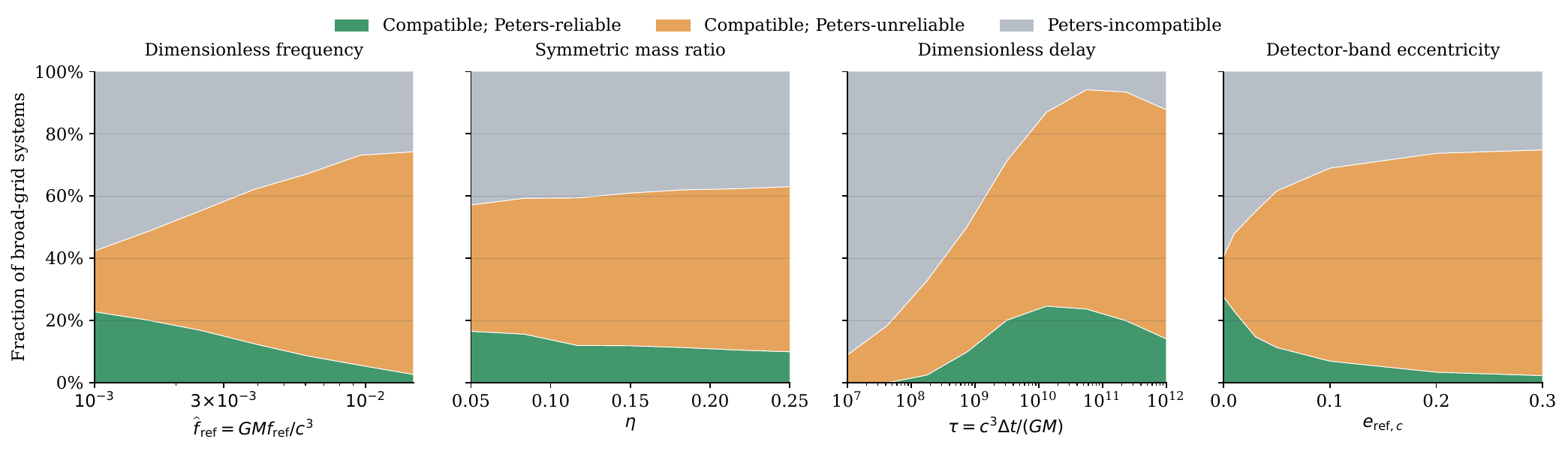}
    \caption{Composition of the broad constraint-atlas grid as a function of
    $\widehat f_{\rm ref}$, $\eta$, $\tau$, and $e_{\rm ref,c}$. Green denotes
    recoil-compatible systems satisfying $R_{\rm sec}\geq0.85$, and orange
    denotes recoil-compatible systems below this threshold. Grey regions
    denote systems for which no compatible progenitor is found in
    the Peters-based scan.}
    \label{fig:atlas_outcome_fractions}
\end{figure*}

We first identify the region of intrinsic parameter space in which the Peters-based construction yields recoil-compatible progenitors and satisfies the reliability criteria introduced in Sec.~\ref{sec:peters_validity}. For this we use the complete broad atlas described in Sec.~\ref{sec:atlas_robustness}, comprising 5733 distinct systems. Of these systems, 3464 ($60.4\%$) have a nonempty recoil-compatible progenitor region. A total of 732 systems satisfy $R_{\rm sec}\geq0.85$, corresponding to $12.8\%$ of the full broad grid and $21.1\%$ of the recoil-compatible systems. Of the 3464 recoil-compatible systems, 2732 therefore fail the Peters-reliability threshold. The other 2269 systems in the broad grid are Peters-incompatible, i.e.~no compatible progenitor is found in the Peters-based scan. Fig.~\ref{fig:atlas_outcome_fractions} shows how these outcomes depend on the four system parameters. These results describe the composition of the sampled grid rather than the expected occurrence rates in an astrophysical population.

An approximate boundary of the Peters-reliable region can be understood by considering post-first-merger target orbits with high initial eccentricity, $e_{\rm post}\simeq1$, that subsequently reach the reference frequency with small but nonzero eccentricity $e_{\rm ref}$. To estimate where these targets reach the orbital-timescale applicability threshold, $N_{\rm orb}=10$, we define $r_g=GM/c^2$, $\bar a=a_{\rm post}/r_g$, and $\bar p=\bar a(1-e_{\rm post}^2)$. The high-eccentricity limit of the merger-time expression in Eq.~\eqref{eq:peters_merger_time},
\begin{equation}
    t_{\rm m}\simeq
    \frac{3}{85}\frac{c^5}{G^3\eta M^3}\,
    a_{\rm post}^4(1-e_{\rm post}^2)^{7/2},
\end{equation}
reduces to
\begin{equation}
    \tau\simeq\frac{3}{85\eta}\,
    \bar a^4(1-e_{\rm post}^2)^{7/2}
    =\frac{3}{85\eta}\,\bar p^{7/2}\bar a^{1/2}.
    \label{eq:atlas_high_e_time}
\end{equation}
The high-eccentricity limit of the Peters invariant in Eq.~\eqref{eq:peters_c0}, given by
\begin{equation}
    c_0 \simeq a_{\rm post}(1-e_{\rm post}^2)
    \left(\frac{425}{304}\right)^{-870/2299},
\end{equation}
can be written as
\begin{equation}
    \bar p\simeq \bar c_0 \left(\frac{425}{304}\right)^{870/2299},
    \label{eq:atlas_high_e_peters_invariant}
\end{equation}
where we define $\bar c_0=c_0/r_g$.

For a target with high initial eccentricity $e_{\rm post}$ that subsequently reaches $f_{\rm ref}$ with small but nonzero eccentricity $e_{\rm ref}$, the small-eccentricity limit of the peak-frequency relation in Eq.~\eqref{eq:wen_frequency_along_peters},
\begin{equation}
    f_{\rm ref} \simeq \frac{\sqrt{GM}}{\pi c_0^{3/2}} e_{\rm ref}^{-18/19},
\end{equation}
implies
\begin{equation}
    \bar c_0\simeq
    (\pi\widehat f_{\rm ref})^{-2/3}e_{\rm ref}^{-12/19}.
    \label{eq:atlas_small_e_invariant}
\end{equation}
Eliminating $\bar a$ from Eq.~\eqref{eq:atlas_high_e_time} yields
\begin{align}
    N_{\rm orb} &= \frac{\Delta t}{2 \pi} \sqrt{\frac{GM}{a_{\rm post}^3}}
    =\frac{\tau}{2\pi\bar a^{3/2}}
      \simeq\frac{1}{2\pi}
      \left(\frac{3}{85}\right)^3
      \frac{\bar p^{21/2}}{\eta^3\tau^2}
      \nonumber\\
    &\simeq
      \frac{1}{2\pi^8} \left(\frac{3}{85}\right)^3 \left(\frac{425}{304}\right)^{9135/2299} \widehat f_{\rm ref}^{-7}\eta^{-3}\tau^{-2}e_{\rm ref}^{-126/19}.
    \label{eq:atlas_orbital_timescale_scaling}
\end{align}
Setting $N_{\rm orb}=10$ therefore defines the approximate critical detector-band eccentricity for the orbital-timescale applicability condition
\begin{equation}
    e_{\rm ref,\rm crit}\simeq{}0.0209
    \left(\frac{\widehat f_{\rm ref}}{0.005}\right)^{-\frac{19}{18}}
    \left(\frac{\eta}{0.25}\right)^{-\frac{19}{42}}
    \left(\frac{\tau}{10^{10}}\right)^{-\frac{19}{63}}.
    \label{eq:atlas_critical_eccentricity}
\end{equation}
Across the sampled atlas, the orbital-timescale condition is violated before the pericenter condition as the target eccentricity increases, so this criterion provides an approximate boundary of the Peters-reliable region. For a detector-band eccentricity interval $\mathcal{E}_{\rm ref}$, the condition $\max\mathcal{E}_{\rm ref}\lesssim e_{{\rm ref},{\rm crit}}$ estimates when the entire admitted target family satisfies the orbital-timescale condition.

We assess the accuracy of this approximation by numerically locating the $N_{\rm orb}=10$ boundary using the full Peters expressions and comparing the resulting critical eccentricity with Eq.~\eqref{eq:atlas_critical_eccentricity}. For all sampled parameter combinations yielding $e_{\rm ref,crit}^{\rm num}\leq0.03$, the relative difference is below $3.2\%$. Larger deviations are expected when the assumptions $e_{\rm post}\simeq1$ and $e_{\rm ref}\ll1$ are not satisfied.
 
We also test this analytical criterion on the recoil-compatible systems in the broad atlas. It correctly identifies 620 of the 732 Peters-reliable systems ($84.7\%$), while all of the 2732 Peters-unreliable systems were correctly identified as unreliable. Of the 112 reliable systems missed, 58 satisfy both applicability conditions for the target orbit corresponding to $\max\mathcal E_{\rm ref}$, evaluated using the full Peters expressions, but fail the approximate analytical criterion. For the remaining 54 systems, this target fails at least one applicability condition, yet reliable targets retain sufficient progenitor volume to satisfy $R_{\rm sec}\geq0.85$. The analytical criterion therefore provides a simple, conservative predictor of Peters reliability among recoil-compatible systems on the sampled grid.

The analytic expression in Eq.~\eqref{eq:atlas_critical_eccentricity} also helps explain the trends in Fig.~\ref{fig:atlas_outcome_fractions}. The critical eccentricity decreases strongly with $\widehat f_{\rm ref}$ and more weakly with $\eta$ and $\tau$, while increasing $e_{{\rm ref},c}$ shifts the admitted eccentricity interval toward or beyond this boundary. These dependencies help explain the decline in the Peters-reliable fraction with increasing reference frequency, symmetric mass ratio, and detector-band eccentricity, whereas the delay dependence also reflects changes in recoil compatibility. 

The first two panels of Fig.~\ref{fig:atlas_outcome_fractions} show that larger $\widehat f_{\rm ref}$ and larger $\eta$ increase the fraction of recoil-compatible systems while decreasing the fraction that is also Peters-reliable. Across the sampled frequency range, compatibility rises from $42.4\%$ to $74.2\%$, whereas the reliable fraction of all systems falls from $22.8\%$ to $2.7\%$. The dependence on $\eta$ is weaker: compatibility increases from $57.1\%$ at $\eta=0.05$ to $63.0\%$ at $\eta=0.25$, while the reliable fraction decreases from $16.6\%$ to $10.0\%$. 

The delay dependence displays the clearest separation between compatibility and reliability. The fraction with no compatible progenitor falls from $91.1\%$ at $\tau=10^7$ to a minimum of $5.8\%$ at $\tau\simeq5.6\times10^{10}$, before increasing to $12.2\%$ at $\tau=10^{12}$. No reliable system is found at either of the two shortest sampled delays, $\tau=10^7$ and $4.2\times10^7$. At longer delays the reliable fraction increases, reaches $24.6\%$ near $\tau=1.3\times10^{10}$, and declines to $14.1\%$ at $\tau=10^{12}$. Longer delays allow for larger target orbits, which can increase the overlap between their accessible radial ranges and the physically admissible progenitor domain. However, a long physical lifetime of the post-first-merger binary does not necessarily imply evolution over many initial orbital periods.

The eccentricity panel shows a complementary competition. The reliable fraction is largest at $e_{\rm ref,c}=0$, where compatibility is least common: $40.7\%$ of systems are compatible and $27.6\%$ of all systems are Peters-reliable. At $e_{\rm ref,c}=0.3$, these fractions become $74.8\%$ and $2.3\%$, respectively. At fixed masses and delay, larger values of $e_{\rm ref,c}$ select larger initial target-orbit eccentricities $e_{\rm post}$. The resulting larger radial extent of the post-first-merger target orbits can improve kinematic accessibility, while the increased initial orbital period and the corresponding smaller value of $N_{\rm orb}$ make orbit-averaged evolution less reliable. Given the small fraction of Peters-reliable systems at $e_{\rm ref,c}=0.3$, we do not extend the constraint atlas to higher values of $e_{\rm ref,c}$.

All reported fractions depend on the sampled parameter domain, which is discussed further in the next section.

\subsubsection{Full constraint atlas}

In this section, we present the full constraint atlas and examine how the constraint-strength quantities vary with the intrinsic system parameters across the sampled parameter space. To construct the full atlas, we augment the broad intrinsic grid with two successive adaptive midpoint-refinement passes within the Peters-reliable regions. Refinement can occur along all four continuous intrinsic coordinates, using geometric midpoints in $\widehat f_{\rm ref}$ and $\tau$ and arithmetic midpoints in $\eta$ and $e_{\rm ref,c}$, with the remaining coordinates and mass-ordering branch held fixed for each selected edge. The passes target reliability and constraint-class transitions, as well as rapid changes in the continuous quantities, providing denser sampling in regions where the inferred constraints change appreciably. The first refinement pass adds 3870 distinct physical systems and the second one adds 9527. Together with the broad atlas grid comprising 5733 systems, the full atlas therefore consists of 19130 systems. Only refinement systems satisfying the numerical convergence criteria defined in Sec.~\ref{sec:convergence_tolerances} have been included.

\begin{figure*}[htp!]
    \centering
    \vspace{-10pt}
    \includegraphics[width=1.0\textwidth]{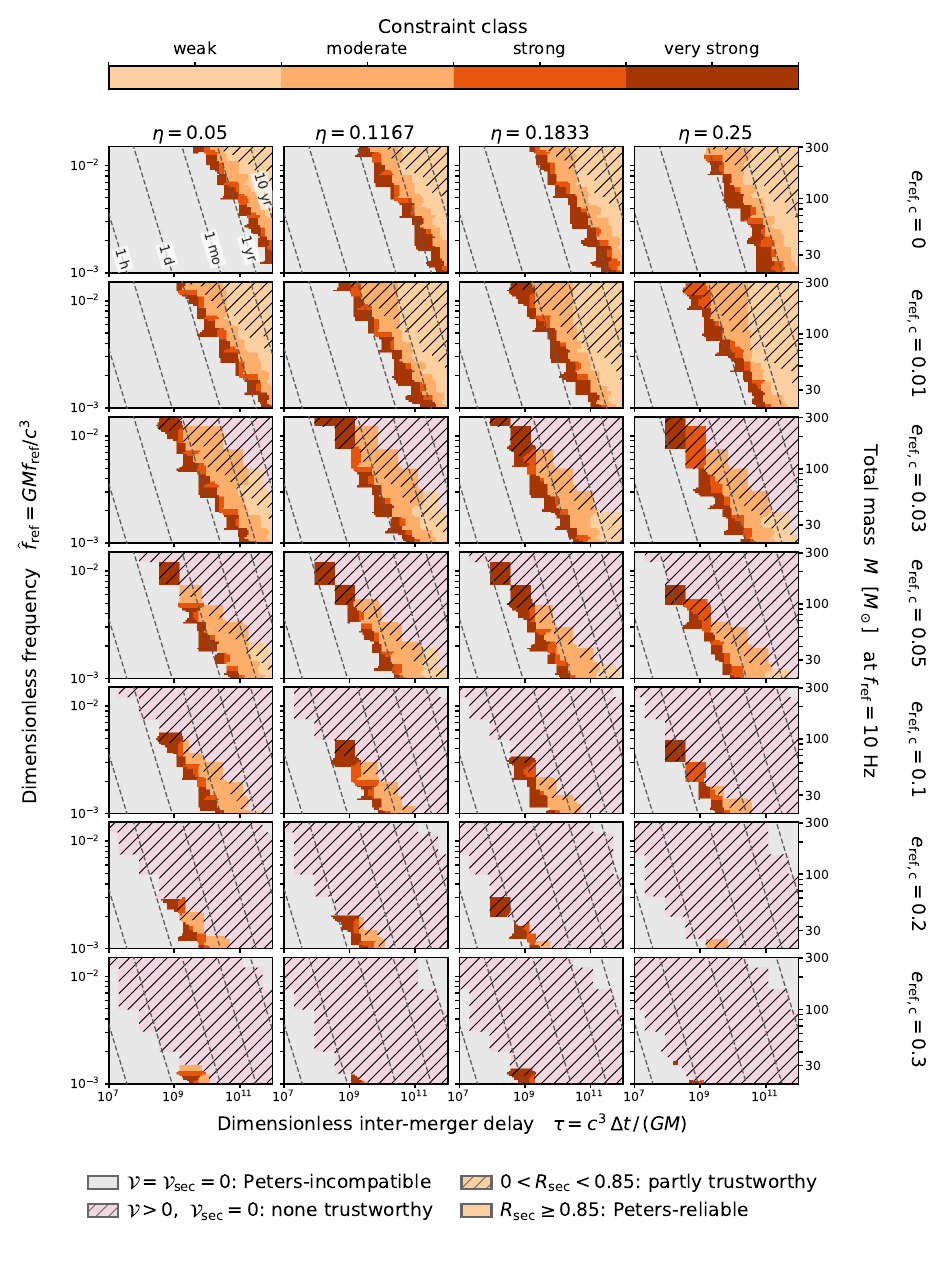}
    \vspace{-30pt}
    \caption{Constraint classes for the mass ordering branch $m_{\rm rem}>m_3$ and their dependence on the dimensionless frequency $\widehat f_{\rm ref}$ (or total mass), inter-merger delay $\tau$, symmetric mass ratio $\eta$, and eccentricity at the reference frequency $e_{\rm ref,c}$.}
    \label{fig:atlas_class}
\end{figure*}

\begin{figure*}[htp!]
    \centering
    \vspace{-10pt}
    \includegraphics[width=1.0\textwidth]{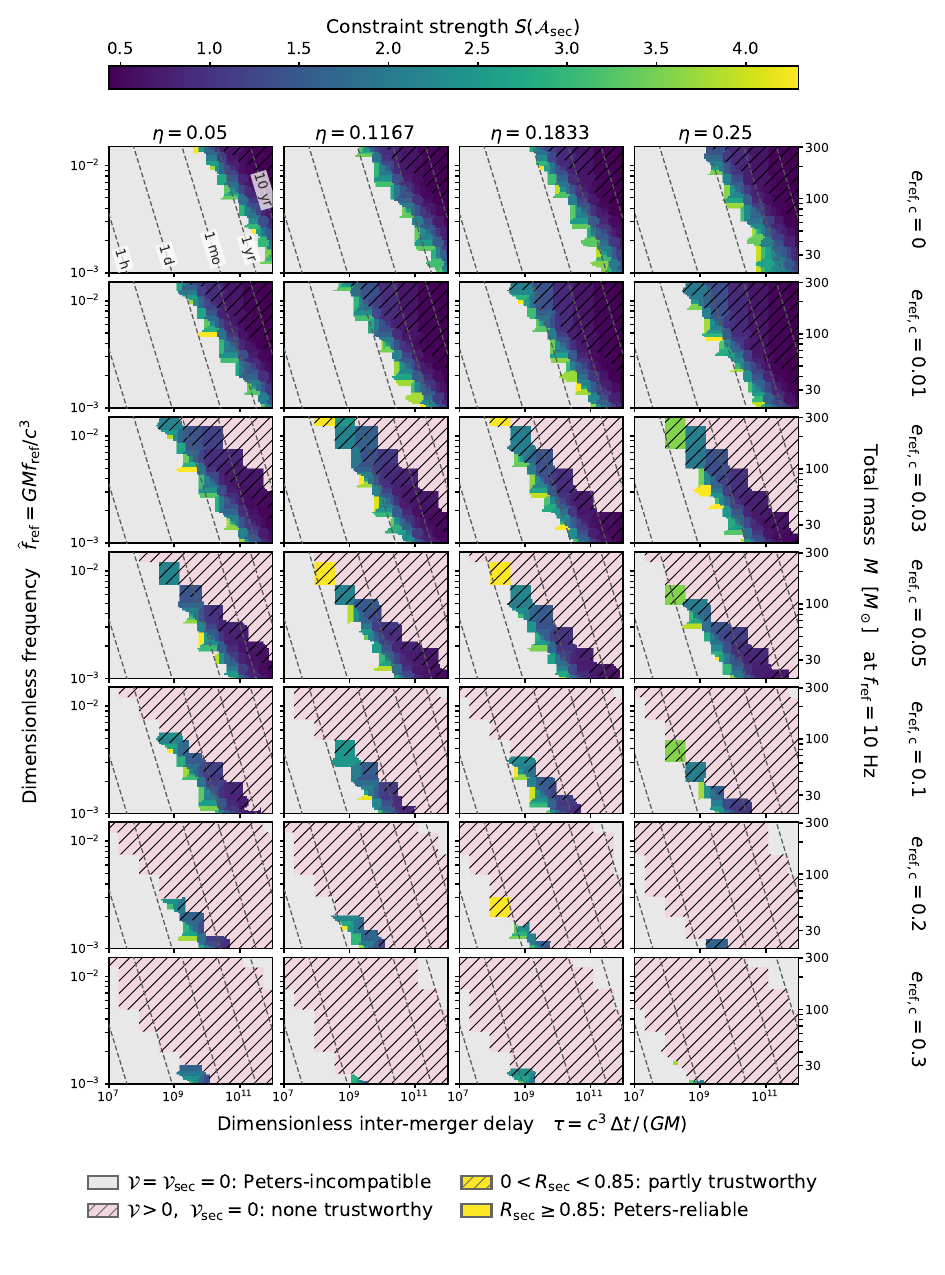}
    \vspace{-30pt}
    \caption{Constraint strengths for the mass ordering branch $m_{\rm rem}>m_3$ and their dependence on the dimensionless frequency $\widehat f_{\rm ref}$ (or total mass), inter-merger delay $\tau$, symmetric mass ratio $\eta$, and eccentricity at the reference frequency $e_{\rm ref,c}$.}
    \label{fig:atlas_S}
\end{figure*}

\begin{figure*}[htp!]
    \centering
    \vspace{-10pt}
    \includegraphics[width=1.0\textwidth]{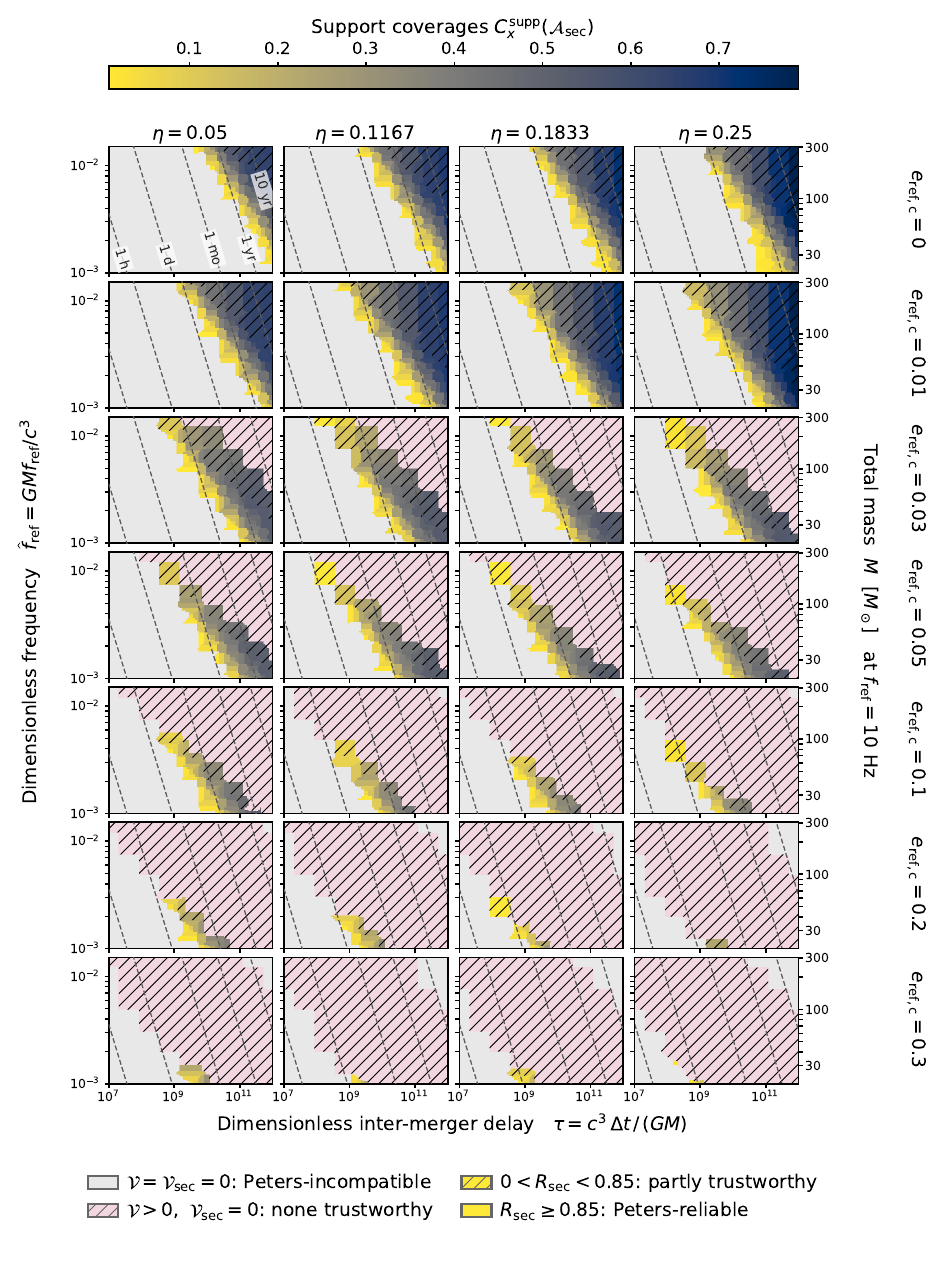}
    \vspace{-30pt}
    \caption{Semi-major axis support coverages for the mass ordering branch $m_{\rm rem}>m_3$ and their dependence on the dimensionless frequency $\widehat f_{\rm ref}$ (or total mass), inter-merger delay $\tau$, symmetric mass ratio $\eta$, and eccentricity at the reference frequency $e_{\rm ref,c}$.}
    \label{fig:atlas_Cx}
\end{figure*}

\begin{figure*}[htp!]
    \centering
    \vspace{-10pt}
    \includegraphics[width=1.0\textwidth]{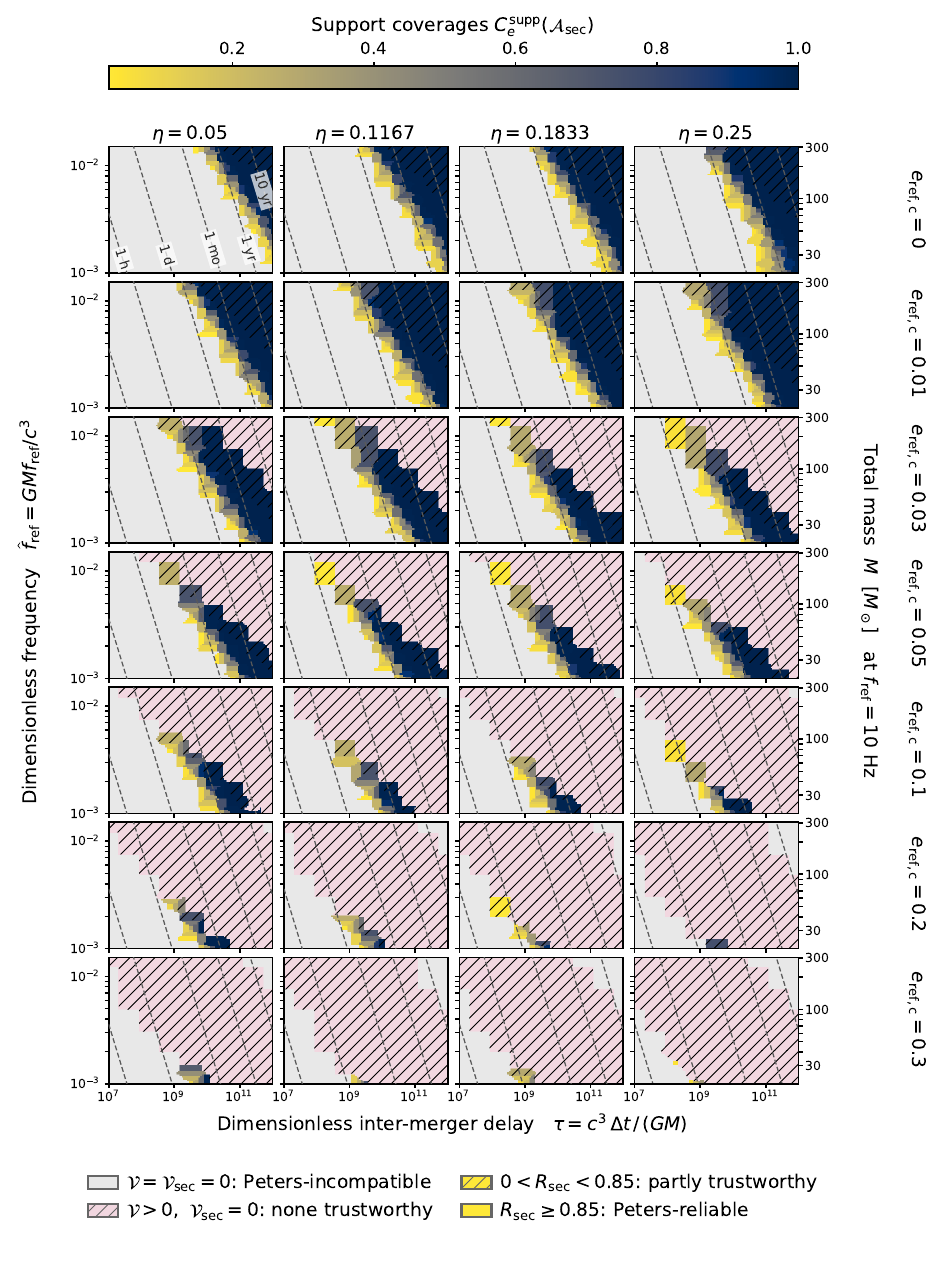}
    \vspace{-30pt}
    \caption{Eccentricity support coverages for the mass ordering branch $m_{\rm rem}>m_3$ and their dependence on the dimensionless frequency $\widehat f_{\rm ref}$ (or total mass), inter-merger delay $\tau$, symmetric mass ratio $\eta$, and eccentricity at the reference frequency $e_{\rm ref,c}$.}
    \label{fig:atlas_Ce}
\end{figure*}

\begin{figure*}[htp!]
    \centering
    \vspace{-10pt}
    \includegraphics[width=1.0\textwidth]{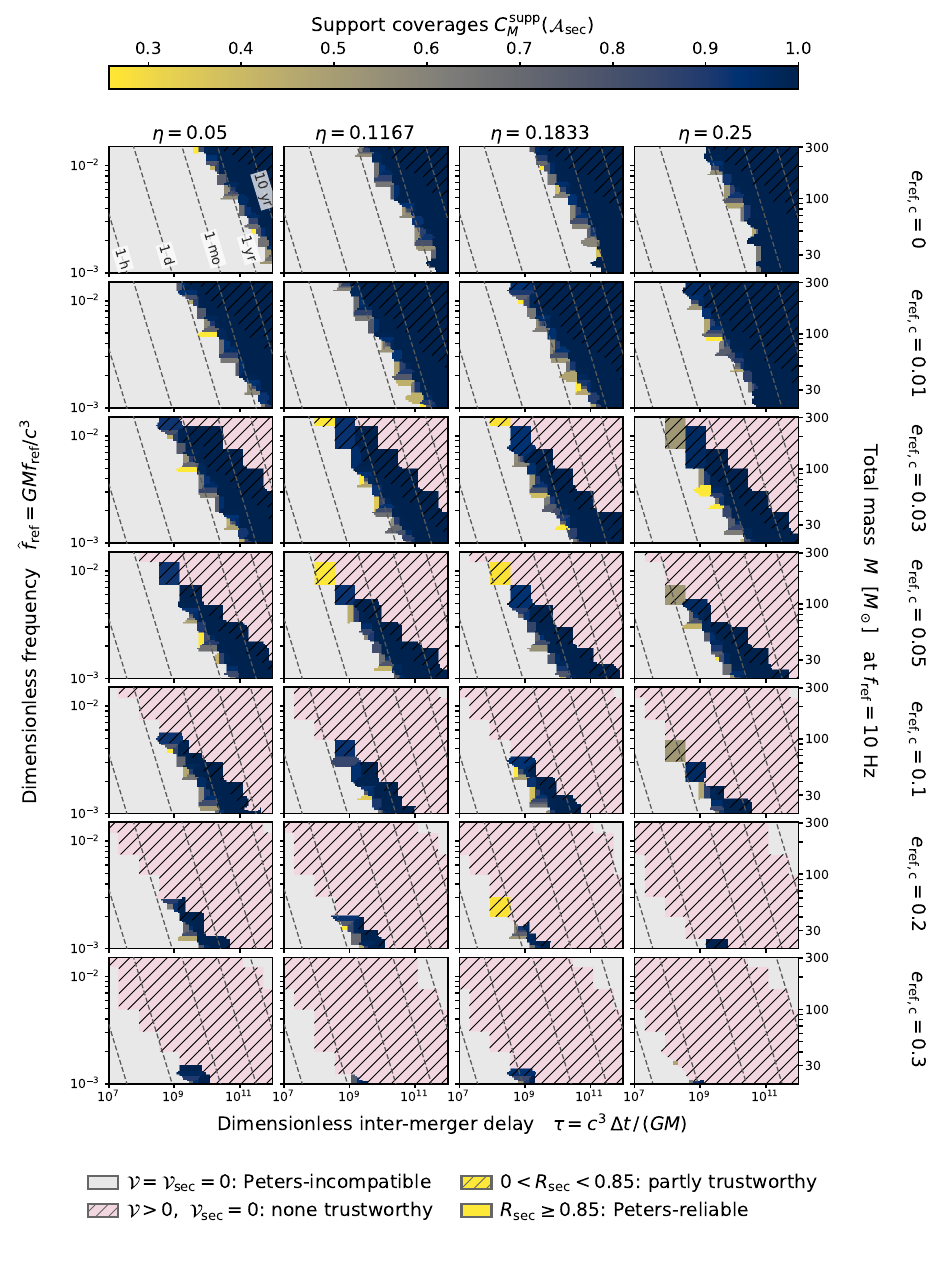}
    \vspace{-30pt}
    \caption{Mean anomaly support coverages for the mass ordering branch $m_{\rm rem}>m_3$ and their dependence on the dimensionless frequency $\widehat f_{\rm ref}$ (or total mass), inter-merger delay $\tau$, symmetric mass ratio $\eta$, and eccentricity at the reference frequency $e_{\rm ref,c}$.}
    \label{fig:atlas_CM}
\end{figure*}

\begin{figure*}[htp!]
    \centering
    \vspace{-10pt}
    \includegraphics[width=1.0\textwidth]{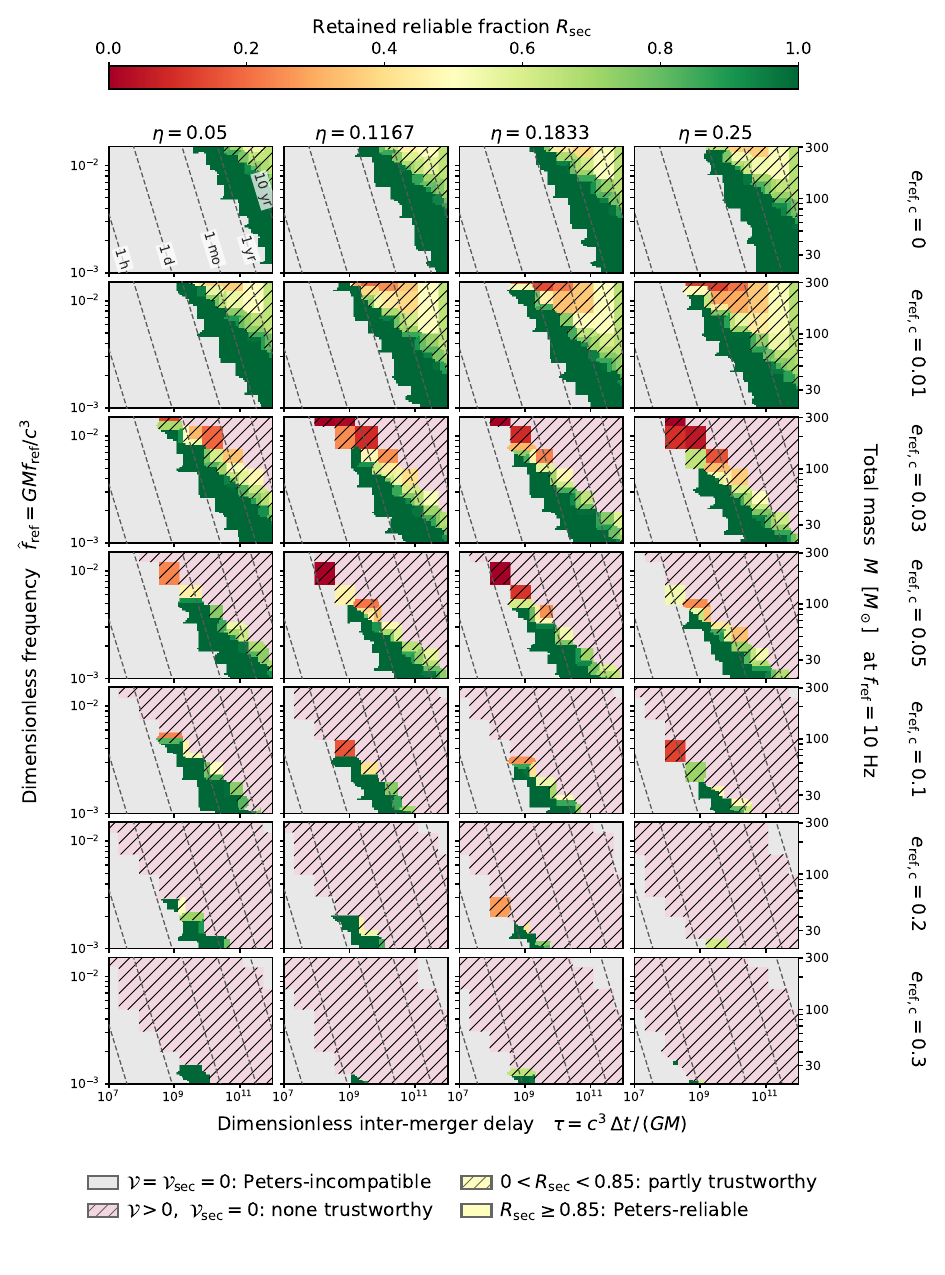}
    \vspace{-30pt}
    \caption{Retained reliable fractions for the mass ordering branch $m_{\rm rem}>m_3$ and their dependence on the dimensionless frequency $\widehat f_{\rm ref}$ (or total mass), inter-merger delay $\tau$, symmetric mass ratio $\eta$, and eccentricity at the reference frequency $e_{\rm ref,c}$.}
    \label{fig:atlas_Rsec}
\end{figure*}

Figures \ref{fig:atlas_class}--\ref{fig:atlas_Rsec} show slices of the full constraint atlas for the mass-ordering branch $m_{\rm rem}>m_3$, depicting the dependence of $R_{\rm sec}$, $S(\mathcal A_{\rm sec})$, $C_X^{\rm supp}(\mathcal A_{\rm sec})$, and the associated constraint classes on the system parameters. Columns correspond to different symmetric mass ratios $\eta$, while rows correspond to different $e_{\rm ref,c}$. Within each panel, the dimensionless inter-merger delay $\tau$ and reference frequency $\widehat f_{\rm ref}$ are shown. The dimensionless axes allow the results to be interpreted without selecting a particular total mass or reference frequency. For orientation, the secondary vertical axis gives the total mass corresponding to a reference frequency of 10 Hz. The dashed lines indicate constant physical inter-merger delays at the same reference frequency. The other mass-ordering branch, $m_3>m_{\rm rem}$, generally exhibits similar qualitative trends and comparable constraint strengths, with a few exceptions that are discussed in Appendix \ref{sec:appendix}.

\begin{figure*}[t]
    \centering
    \includegraphics[width=\textwidth]{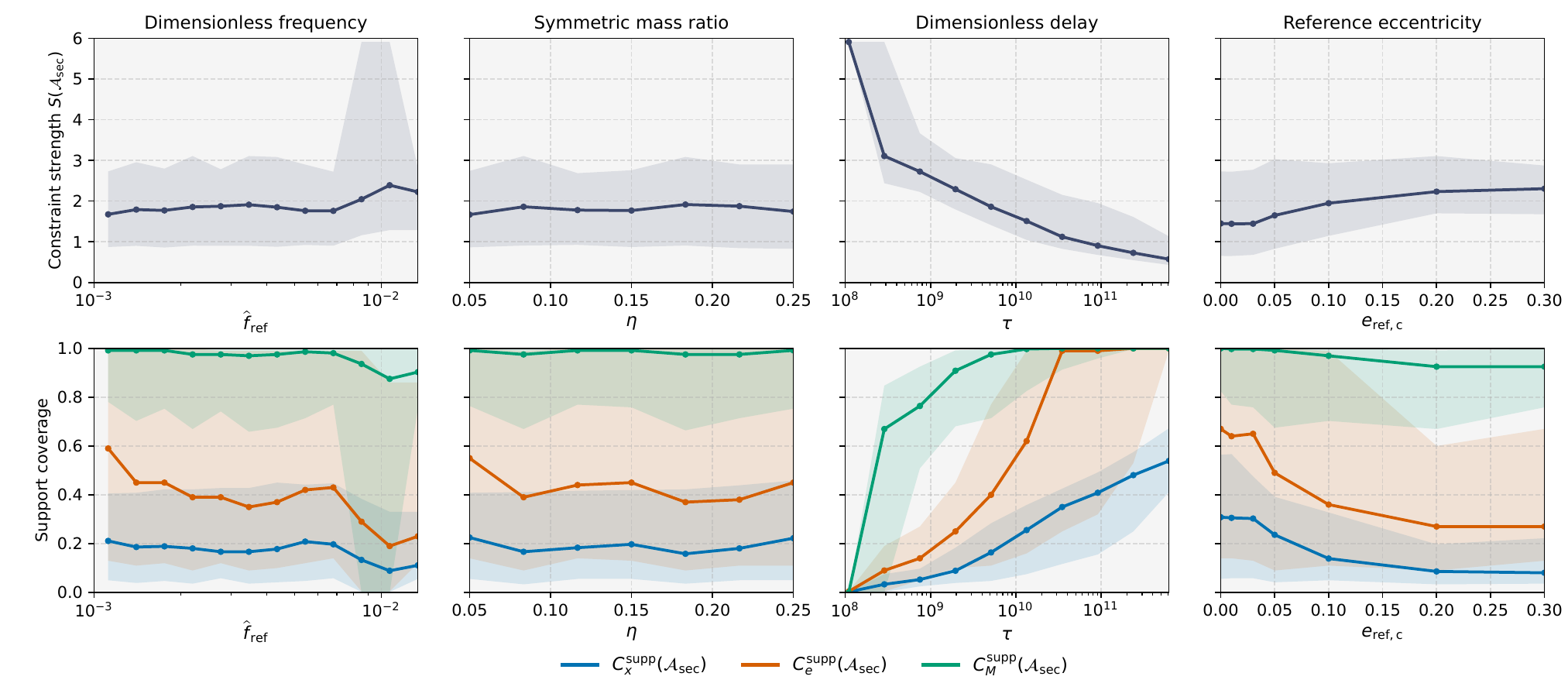}
    \caption{Dependence of the constraint strength $S(\mathcal A_{\rm sec})$ (top row) and support coverages $C_X^{\rm supp}(\mathcal A_{\rm sec})$, with $X\in \{x,e,M\}$ (bottom row), on the system parameters for the mass-ordering branch $m_{\rm rem}>m_3$. Only Peters-reliable systems are included. Solid curves show weighted medians over the remaining parameters, and shaded regions indicate the 16th–84th percentiles. }
    \label{fig:atlas_trends}
\end{figure*}

\newpage

The atlas exhibits a narrow band of strong constraints that extends approximately diagonally across the logarithmic panels. Along this band, the allowed recoil-compatible volume fractions are small, producing large values of the constraint strength $S(\mathcal A_{\rm sec})$. Substantial portions of the band lie within the Peters-reliable region. However, strong constraints also occur in partially reliable regions, where the inferred constraint strengths should be interpreted together with the corresponding values of $R_{\rm sec}$ presented in Fig.~\ref{fig:atlas_Rsec}.

The diagonal strong-constraint band is well described by an empirical power law. Fitting the locations of the maxima of $S(\mathcal A_{\rm sec})$ at fixed $\widehat f_{\rm ref}$ within the Peters-reliable region, we obtain
\begin{equation}
\begin{aligned}
    \tau_{\rm band}\simeq {}&
    7.906\times10^8
    \left(\frac{\widehat f_{\rm ref}}{0.005}\right)^{-2.134}
    \left(\frac{\eta}{0.15}\right)^{-0.868}\ \\
    &\times
    \left(\frac{e_{\rm ref,c}+0.01}{0.05}\right)^{-1.968}
    10^{0.140s\sqrt{1-4\eta}},
    \label{eq:strong_constraint_fit}
\end{aligned}
\end{equation}
where $s=+1$ for $m_{\rm rem}>m_3$ and $s=-1$ for the case $m_3>m_{\rm rem}$. The branch-dependent factor approaches unity as $\eta\rightarrow0.25$, where the two mass orderings become equivalent. At fixed eccentricity and symmetric mass ratio, this relation corresponds to $\widehat f_{\rm ref}\propto\tau^{-0.469}$. Increasing either $e_{\rm ref,c}$ or $\eta$ shifts the diagonal band toward shorter dimensionless delays at fixed $\widehat f_{\rm ref}$.

The position and extent of the strong-constraint band are robust under the tested extensions of the reference domain, consistent with our findings in Sec.~\ref{sec:atlas_robustness} that such extensions generally do not affect strongly constrained systems. Much of the band lies within regions where $R_{\rm sec}$ is close to unity, so modest changes to the threshold used to classify systems as Peters-reliable are also expected to leave these portions unaffected. Varying the detector eccentricity window $\mathcal E_{\rm ref}$ or the upper recoil-velocity limit modifies the allowed configurations and can therefore change the constraint strengths and the extent of the strongly constrained region. The position of the band may remain approximately unchanged if these variations affect neighboring systems similarly.

The retained reliable fractions $R_{\rm sec}$ depicted in Fig.~\ref{fig:atlas_Rsec} reveal a systematic transition across the compatible region. Near the strong-constraint band, $R_{\rm sec}$ is close to unity, although the region with $R_{\rm sec}\simeq1$ extends far beyond the band itself. Moving toward larger $\tau$ at fixed $\widehat f_{\rm ref}$, or toward larger $\widehat f_{\rm ref}$ at fixed $\tau$, the retained fraction decreases through a partially reliable transition region, where smaller values of $R_{\rm sec}$ are generally obtained for longer inter-merger delays and higher-mass systems. As $e_{\rm ref,c}$ and $\eta$ increase, the region satisfying $R_{\rm sec}\geq0.85$ becomes increasingly confined to lower $\widehat f_{\rm ref}$, corresponding to lower-mass systems.

The typical inter-merger delays associated with strong constraints depend strongly on the eccentricity of the second merger. At $f_{\rm ref}=10$ Hz, nearly circular second mergers can yield strong constraints for inter-merger delays of several months to a few years. As the reference eccentricity increases, the band moves toward shorter physical delays. Eccentricities of around 0.05 admit characteristic delays of weeks to months, while systems with $e_{\rm ref,c}\gtrsim0.1$ commonly yield strong constraints at delays of order days.

The support coverages reveal that the different outer-orbit parameters are constrained to different degrees. Fig.~\ref{fig:atlas_Cx} shows that the separation coordinate $x$ remains constrained over a broad region of the atlas. Its strongest constraints occur near the narrow strong-constraint band and gradually become weaker away from it. Constraints on the semi-major axis therefore remain useful even outside this narrow band.

The constraints on the eccentricity depicted in Fig.~\ref{fig:atlas_Ce} exhibit much sharper transitions. Near the strong-constraint band, $C_e^{\rm supp}(\mathcal A_{\rm sec})$ can be small, whereas farther away from the band it increases rapidly toward broad coverage of the sampled eccentricity interval. Strong constraints on the outer eccentricity consequently occupy a narrower range of intrinsic parameters than constraints on the semi-major axis.

The outer orbital phase is generally the most weakly constrained coordinate. Its mean-anomaly support coverage, $C_M^{\rm supp}(\mathcal A_{\rm sec})$, depicted in Fig.~\ref{fig:atlas_CM}, remains close to unity over much of the allowed region. Strong constraints on this parameter occur mainly in small localized islands close to the strong-constraint band. Therefore, even systems with a small allowed recoil-compatible volume fraction need not tightly constrain the phase. 

Fig.~\ref{fig:atlas_trends} summarizes how the constraint quantities vary with each intrinsic parameter, showing weighted medians and the 16th–84th percentile ranges over the remaining parameters within the Peters-reliable region. The results show no strong trends with the total mass and symmetric mass ratio of the system, and clearer trends towards stronger constraints for systems with shorter inter-merger delays and higher detector-band eccentricities, although Peters-reliable systems occupy a smaller portion of the sampled parameter space in these regimes. The different behavior of the support coverages can also be seen very clearly, since they approach unity at different rates as $\tau$ increases.

Several aspects of the composition of the broad constraint atlas depicted in Fig.~\ref{fig:atlas_outcome_fractions} are also visible in the full atlas figures. An additional feature that further explains some of the trends is the apparent diagonal-band structure of the compatible region that is especially visible in the red hatched area in the plots corresponding to the two highest eccentricity values $e_{\rm ref,c}=0.2$ and $e_{\rm ref,c}=0.3$. This region looks like a diagonal band in the logarithmic panels with Peters-incompatible regions visible on either side. As $e_{\rm ref,c}$ and $\eta$ increase, the band shifts toward shorter $\tau$, changing its overlap with the sampled parameter domain. At low values of $e_{\rm ref,c}$ and $\eta$ the band is not fully inside the sampled parameter range of $\widehat f_{\rm ref}$ and $\tau$ and gradually moves into the range for increasing values of $e_{\rm ref,c}$ and $\eta$. This displacement helps to explain both the increasing fraction of compatible systems with eccentricity and its subsequent flattening: the overlap of the band with the sampled parameter domain grows initially, but changes less at around $e_{\rm ref,c}\approx0.2$, where the band has fully entered the domain. The maximum of compatible systems can be found at $\tau\approx 10^{11}$, which is the slice that is mostly inside the band of compatible systems. For $\tau> 10^{11}$ there are already Peters-incompatible regions since the band of compatible systems has moved further towards lower values of $\tau$ at high $e_{\rm ref,c}$. This behavior illustrates the dependence of the results of Fig.~\ref{fig:atlas_outcome_fractions} on the sampled parameter domain. If the band keeps shifting towards lower values of $\tau$ with increasing values of $e_{\rm ref,c}$, the fraction of compatible systems should have reached its maximum around $e_{\rm ref,c}=0.3$ and should go down again with increasing $e_{\rm ref,c}$. The relatively flat trends with $\eta$ in Fig.~\ref{fig:atlas_outcome_fractions} are consistent with the fact that dependence on $\eta$ is weaker than that on frequency and eccentricity.

Finally, we consider the interpretation of the Peters-incompatible regions. Towards smaller values of $\tau$, these regions arise as the recoil-compatible volume fraction approaches zero: the constraints progressively tighten until no compatible progenitor configuration remains. When the surviving region satisfies $R_{\rm sec}=1$ up to this transition, its disappearance provides stronger evidence for general physical incompatibility within the assumed hierarchical-triple model. However, this interpretation remains subject to numerical resolution and to the residual errors of the Peters approximation, even if our reliability criteria are satisfied. The Peters-incompatible regions at larger values of $\tau$ have a less straightforward interpretation, since the recoil-compatible volume generally increases and the constraints weaken with increasing $\tau$ over the sampled domain. However, the accompanying decline in $R_{\rm sec}$ makes extrapolation of this trend less reliable. In general, Peters-incompatibility only establishes the absence of compatible configurations within our approximation and for the sampled parameters, rather than conclusively excluding a general hierarchical-triple origin. A stronger exclusion would require demonstrating robustness to parameter uncertainties, the search domain, numerical resolution, and the use of more accurate dynamical treatments.

\subsection{Constraints on selected hierarchical-merger candidates}

In this section, we investigate how well our method constrains the progenitor orbits of existing hierarchical-merger candidates. We consider seven distinct pairs of observed gravitational-wave events proposed in the literature as possible successive mergers in a hierarchical-merger chain \citep{Veske2020, Veske2021}. Candidates identified solely from the properties of a single merger are excluded because applying our method requires an identified candidate predecessor and an inter-merger delay. For GW190514–GW190521, we examine both primary- and secondary-remnant assignments, following the interpretations considered by Refs.~\citep{Veske2021, Li2025}. Where the literature does not specify the matched component, we adopt the component whose representative source-frame mass is closest to the earlier remnant mass.

\begin{table*}[t]
\centering
\newcommand{\candmassbox}[1]{%
  \boxed{\makebox[6em][c]{$\displaystyle
    \vphantom{\begin{gathered}152.4^{+33.6}_{-30.1}\\[2pt]152.4^{+33.6}_{-30.1}\end{gathered}}%
    \smash{#1}$}}}
\caption{Selected candidate pairs of successive mergers. Superscripts (1) and (2) label the first and second observed mergers. For each system, the source-frame masses are given in the upper line and the detector-frame masses in the lower line, both in $M_\odot$ and with $m_1\geq m_2$. The masses are given as posterior medians with 90\% credible intervals. A box encloses the masses for the second-merger component assumed to contain the first remnant (for GW190514–GW190521 we investigate both possibilities).}
\label{tab:selected_merger_candidates}
\small
\setlength{\tabcolsep}{3pt}
\renewcommand{\arraystretch}{1.35}
\resizebox{\textwidth}{!}{%
\begin{tabular}{@{}lccccccc@{}}
\toprule
\shortstack[l]{First merger\\$\to$ second merger} &
$m_1^{(1)}$ & $m_2^{(1)}$ & $m_{\rm rem}^{(1)}$ &
$m_1^{(2)}$ & $m_2^{(2)}$ &
\shortstack{$\Delta t_{\rm obs}$\\(days)} &
\shortstack{Second-merger\\eccentricity} \\
\midrule
\shortstack[l]{GW151012$\to$GW170809} & $\begin{gathered}24.8^{+14.5}_{-6.3}\\[2pt]29.6^{+17.1}_{-6.9}\end{gathered}$ & $\begin{gathered}13.6^{+4.5}_{-4.9}\\[2pt]16.3^{+5.7}_{-6.0}\end{gathered}$ & $\begin{gathered}37.1^{+10.6}_{-4.6}\\[2pt]44.3^{+12.7}_{-4.3}\end{gathered}$ & $\candmassbox{\begin{gathered}34.1^{+8.0}_{-5.3}\\[2pt]41.2^{+9.6}_{-6.3}\end{gathered}}$ & $\begin{gathered}24.2^{+4.8}_{-5.3}\\[2pt]29.2^{+5.6}_{-6.5}\end{gathered}$ & 666.940 & \shortstack{No compelling evidence\\for eccentricity} \\\addlinespace[4pt]
\shortstack[l]{GW190514$\to$GW190521} & $\begin{gathered}40.9^{+17.3}_{-9.3}\\[2pt]67.0^{+21.4}_{-12.6}\end{gathered}$ & $\begin{gathered}28.4^{+10.0}_{-10.1}\\[2pt]47.3^{+12.1}_{-18.2}\end{gathered}$ & $\begin{gathered}66.4^{+19.0}_{-11.5}\\[2pt]109.7^{+18.9}_{-17.6}\end{gathered}$ & $\candmassbox{\begin{gathered}98.4^{+33.6}_{-21.7}\\[2pt]152.4^{+31.6}_{-17.6}\end{gathered}}$ & $\candmassbox{\begin{gathered}57.2^{+27.1}_{-30.1}\\[2pt]89.7^{+49.0}_{-52.2}\end{gathered}}$ & 6.839 & \shortstack{Model-dependent evidence\\for eccentricity} \\\addlinespace[4pt]
\shortstack[l]{GW170729$\to$GW190521} & $\begin{gathered}54.7^{+12.7}_{-12.8}\\[2pt]77.9^{+17.9}_{-14.5}\end{gathered}$ & $\begin{gathered}30.2^{+11.9}_{-10.2}\\[2pt]44.1^{+18.1}_{-17.5}\end{gathered}$ & $\begin{gathered}80.3^{+13.5}_{-10.2}\\[2pt]116.7^{+18.9}_{-18.3}\end{gathered}$ & $\candmassbox{\begin{gathered}98.4^{+33.6}_{-21.7}\\[2pt]152.4^{+31.6}_{-17.6}\end{gathered}}$ & $\begin{gathered}57.2^{+27.1}_{-30.1}\\[2pt]89.7^{+49.0}_{-52.2}\end{gathered}$ & 660.338 & \shortstack{Model-dependent evidence\\for eccentricity} \\\addlinespace[4pt]
\shortstack[l]{GW190708$\to$GW190915} & $\begin{gathered}19.8^{+4.3}_{-4.3}\\[2pt]23.4^{+5.0}_{-5.1}\end{gathered}$ & $\begin{gathered}11.6^{+3.1}_{-2.0}\\[2pt]13.7^{+3.6}_{-2.2}\end{gathered}$ & $\begin{gathered}30.1^{+2.9}_{-2.1}\\[2pt]35.5^{+3.0}_{-1.7}\end{gathered}$ & $\candmassbox{\begin{gathered}32.6^{+8.8}_{-4.9}\\[2pt]43.0^{+11.1}_{-6.0}\end{gathered}}$ & $\begin{gathered}24.5^{+4.9}_{-5.8}\\[2pt]32.5^{+5.6}_{-8.0}\end{gathered}$ & 69.022 & \shortstack{No compelling evidence\\for eccentricity} \\\addlinespace[4pt]
\shortstack[l]{GW190512$\to$GW190910} & $\begin{gathered}23.2^{+5.6}_{-5.6}\\[2pt]29.3^{+6.9}_{-6.7}\end{gathered}$ & $\begin{gathered}12.5^{+3.5}_{-2.6}\\[2pt]15.8^{+4.6}_{-3.0}\end{gathered}$ & $\begin{gathered}34.3^{+4.1}_{-3.4}\\[2pt]43.4^{+4.4}_{-2.7}\end{gathered}$ & $\begin{gathered}43.8^{+7.6}_{-6.8}\\[2pt]56.5^{+8.7}_{-6.5}\end{gathered}$ & $\candmassbox{\begin{gathered}34.2^{+6.6}_{-7.3}\\[2pt]44.5^{+7.5}_{-9.2}\end{gathered}}$ & 120.723 & \shortstack{No compelling evidence\\for eccentricity} \\\addlinespace[4pt]
\shortstack[l]{GW170818$\to$GW190519} & $\begin{gathered}34.8^{+6.5}_{-4.2}\\[2pt]42.1^{+8.0}_{-5.0}\end{gathered}$ & $\begin{gathered}27.6^{+4.1}_{-5.1}\\[2pt]33.4^{+5.1}_{-6.1}\end{gathered}$ & $\begin{gathered}59.7^{+4.9}_{-4.2}\\[2pt]72.3^{+5.9}_{-5.2}\end{gathered}$ & $\candmassbox{\begin{gathered}65.1^{+10.8}_{-11.0}\\[2pt]94.8^{+15.8}_{-12.4}\end{gathered}}$ & $\begin{gathered}40.8^{+11.5}_{-12.7}\\[2pt]59.9^{+16.9}_{-18.9}\end{gathered}$ & 639.549 & \shortstack{No compelling evidence\\for eccentricity} \\\addlinespace[4pt]
\shortstack[l]{GW170729$\to$GW170817A} & $\begin{gathered}54.7^{+12.7}_{-12.8}\\[2pt]77.9^{+17.9}_{-14.5}\end{gathered}$ & $\begin{gathered}30.2^{+11.9}_{-10.2}\\[2pt]44.1^{+18.1}_{-17.5}\end{gathered}$ & $\begin{gathered}80.3^{+13.5}_{-10.2}\\[2pt]116.7^{+18.9}_{-18.3}\end{gathered}$ & $\candmassbox{\begin{gathered}56^{+16}_{-10}\,\\[2pt]88.5^{+15.0}_{-10.8}\end{gathered}}$ & $\begin{gathered}40^{+10}_{-11}\,\\[2pt]71.5^{+11.0}_{-14.0}\end{gathered}$ & 18.338 & \shortstack{No published\\constraints found}\\\addlinespace[4pt]
\bottomrule
\end{tabular}%
}
\end{table*}

Table~\ref{tab:selected_merger_candidates} summarizes the selected pairs, their source-frame and detector-frame masses, detector-frame inter-merger delays, and the available information on the second merger's eccentricity. Source-frame masses are quoted from the literature \citep{GWTC21, Zackay2021}, while detector-frame masses and their uncertainties are derived from publicly released posterior samples. The observer-frame delays $\Delta t_{\rm obs}$ are calculated from reported event GPS times. Eccentricity classifications summarize the available literature. Four of the six distinct second mergers show no compelling evidence for eccentricity, with reported upper bounds depending on the waveform model and prior assumptions \citep{RomeroShaw2019,RomeroShaw2022,Gupte2025}. For GW190521, published interpretations range from quasi-circular mergers to substantially eccentric mergers and even a hyperbolic capture scenario \citep{RomeroShaw2020,Gayathri2022,Gamba2023,RamosBuades2023,Gupte2025}. For GW170817A, we found no published eccentricity constraints.

\begin{table*}[t]
\centering
\caption{Constraints for the selected candidate merger pairs,
using detector-frame median masses and the observed inter-merger delays
reported in Table~\ref{tab:selected_merger_candidates}. The imposed
second-merger eccentricity upper limits refer to an adopted reference
frequency of $10\,\mathrm{Hz}$.
For GW190514$\to$GW190521, (1) and (2) denote primary- and secondary-remnant
assignments. Constraint classes and strengths refer to the
secular-compatible region. Semi-major axes are detector-frame values,
with $a_{\rm out}^{\rm det}=(1+z)a_{\rm out}^{\rm src}$, and
$\mathcal M_{\rm m}$ is the mean anomaly at the first merger.
Asterisks mark endpoints at the limits of the search domain rather than
inferred strict bounds. Boundary values are rounded
to the decimal place of the leading significant digit of the associated uncertainty.}
\label{tab:selected_merger_constraints}
\small
\setlength{\tabcolsep}{4pt}
\renewcommand{\arraystretch}{1.15}
\resizebox{\textwidth}{!}{%
\begin{tabular}{@{}lccccccc@{}}
\toprule
\shortstack[l]{First merger\\$\to$ second merger} &
\shortstack{Second-merger\\eccentricity $e_{10}$} &
Constraint class & $S(\mathcal A_{\rm sec})$ &
$R_{\rm sec}$ &
\shortstack{Constraints for $a_{\rm out}^{\rm det}$\\(AU)} &
\shortstack{Constraints for $e_{\rm out}$} &
\shortstack{Constraints for $\mathcal M_{\rm m}$\\(deg)} \\
\midrule
GW151012$\to$GW170809
& $e_{10}<0.01$ & Weak & $0.63$ & $1.000$
& $[0.00728,\,10^{*}]$
& $[0.000,\,0.999^{*}]$ & $[-180,\,180]$ \\
& $e_{10}<0.02$ & Weak & $0.43$ & $0.683$
& $[0.00728,\,10^{*}]$
& $[0.000,\,0.999^{*}]$ & $[-180,\,180]$ \\
& $e_{10}<0.05$ & Weak & $0.43$ & $0.682$
& $[0.00728,\,10^{*}]$
& $[0.000,\,0.999^{*}]$ & $[-180,\,180]$ \\
\addlinespace[4pt]
GW190514$\to$GW190521 (1)
& $e_{10}<0.01$ & Peters-incompatible & --- & ---
& $\varnothing$ & $\varnothing$ & $\varnothing$ \\
& $e_{10}<0.02$ & Very strong & $2.99$ & $1.000$
& $[0.0308,\,0.0548]$
& $[0.838,\,0.924]$ & $59.33\leq|\mathcal M_{\rm m}|\leq180$ \\
& $e_{10}<0.05$ & Strong & $1.99$ & $0.033$
& $[0.03107,\,0.1807]$
& $[0.719,\,0.972]$ & $12.4\leq|\mathcal M_{\rm m}|\leq180$ \\
\addlinespace[4pt]
GW190514$\to$GW190521 (2)
& $e_{10}<0.01$ & Peters-incompatible & --- & ---
& $\varnothing$ & $\varnothing$ & $\varnothing$ \\
& $e_{10}<0.02$ & Very strong & $2.11$ & $0.777$
& $[0.0373,\,0.1731]$
& $[0.761,\,0.967]$ & $16.7\leq|\mathcal M_{\rm m}|\leq180$ \\
& $e_{10}<0.05$ & Very strong & $2.11$ & $0.019$
& $[0.0373,\,0.1731]$
& $[0.761,\,0.967]$ & $16.7\leq|\mathcal M_{\rm m}|\leq180$ \\
\addlinespace[4pt]
GW170729$\to$GW190521
& $e_{10}<0.01$ & Weak & $0.49$ & $0.681$
& $[0.0237,\,10^{*}]$
& $[0.000,\,0.999^{*}]$ & $[-180,\,180]$ \\
& $e_{10}<0.02$ & Weak & $0.49$ & $0.681$
& $[0.0237,\,10^{*}]$
& $[0.000,\,0.999^{*}]$ & $[-180,\,180]$ \\
& $e_{10}<0.05$ & Weak & $0.49$ & $0.681$
& $[0.0237,\,10^{*}]$
& $[0.000,\,0.999^{*}]$ & $[-180,\,180]$ \\
\addlinespace[4pt]
GW190708$\to$GW190915
& $e_{10}<0.01$ & Peters-incompatible & --- & ---
& $\varnothing$ & $\varnothing$ & $\varnothing$ \\
& $e_{10}<0.02$ & Moderate & $1.29$ & $1.000$
& $[0.00843,\,0.785]$
& $[0.291,\,0.999^{*}]$ & $[-180,\,180]$ \\
& $e_{10}<0.05$ & Weak & $0.74$ & $0.381$
& $[0.00843,\,10^{*}]$
& $[0.000,\,0.999^{*}]$ & $[-180,\,180]$ \\
\addlinespace[4pt]
GW190512$\to$GW190910
& $e_{10}<0.01$ & Strong & $1.73$ & $1.000$
& $[0.01102,\,0.1710]$
& $[0.645,\,0.989]$ & $3.7\leq|\mathcal M_{\rm m}|\leq180$ \\
& $e_{10}<0.02$ & Weak & $0.69$ & $0.720$
& $[0.01102,\,10^{*}]$
& $[0.000,\,0.999^{*}]$ & $[-180,\,180]$ \\
& $e_{10}<0.05$ & Weak & $0.69$ & $0.403$
& $[0.01102,\,10^{*}]$
& $[0.000,\,0.999^{*}]$ & $[-180,\,180]$ \\
\addlinespace[4pt]
GW170818$\to$GW190519
& $e_{10}<0.01$ & Weak & $0.47$ & $0.773$
& $[0.0139,\,10^{*}]$
& $[0.000,\,0.999^{*}]$ & $[-180,\,180]$ \\
& $e_{10}<0.02$ & Weak & $0.47$ & $0.675$
& $[0.0139,\,10^{*}]$
& $[0.000,\,0.999^{*}]$ & $[-180,\,180]$ \\
& $e_{10}<0.05$ & Weak & $0.47$ & $0.675$
& $[0.0139,\,10^{*}]$
& $[0.000,\,0.999^{*}]$ & $[-180,\,180]$ \\
\addlinespace[4pt]
GW170729$\to$GW170817A
& $e_{10}<0.01$ & Peters-incompatible & --- & ---
& $\varnothing$ & $\varnothing$ & $\varnothing$ \\
& $e_{10}<0.02$ & Moderate & $1.46$ & $0.741$
& $[0.02945,\,0.986]$
& $[0.477,\,0.998]$ & $0.5\leq|\mathcal M_{\rm m}|\leq180$ \\
& $e_{10}<0.05$ & Moderate & $1.46$ & $0.074$
& $[0.02950,\,0.979]$
& $[0.477,\,0.997]$ & $0.4\leq|\mathcal M_{\rm m}|\leq180$ \\
\bottomrule
\end{tabular}%
}
\end{table*}

With these data, we first perform a preliminary constraint-strength analysis using only the posterior median detector-frame masses to estimate the attainable constraint strengths and identify promising systems for more detailed subsequent analyses incorporating the full posterior distributions. For each candidate pair, we associate one component of the second merger with the first remnant as described earlier. We then adopt the reported remnant mass of the first merger as $m_{\rm rem}$ and the mass of the other component of the second merger as the tertiary mass. The associated second-merger component mass is therefore not used in these preliminary scans. The dynamical calculations use masses and times in the detector frame. This choice is convenient because the inter-merger delay is measured to high precision in the detector frame and avoids introducing a redshift conversion into the source frame. Because the eccentricity is not well constrained for all second mergers and since all of them allow for a quasi-circular interpretation, we examine three illustrative upper limits on the second-merger eccentricity, $e_{10}<0.01$, $e_{10}<0.02$, and $e_{10}<0.05$, at a reference frequency of $10\,\mathrm{Hz}$. Here, $e_{10}$ is defined through the Peters evolution and the adopted peak-frequency prescription in Eq.~\eqref{eq:wen_peak_frequency}. Its comparison with waveform-based observational constraints requires matching the eccentricity definition and reference-frequency convention \citep{Shaikh2023}. Each scan uses the resolutions, refinement procedure, and convergence criteria described in Sec.~\ref{sec:convergence_tolerances}. We use a restricted reference domain with $a^{\rm det}_{\rm out} \in [0.001, 10]$ AU, $e_{\rm out} \in [0.0, 0.999]$, and $\mathcal M_{\rm m} \in [-180^\circ, 180^\circ]$. For this domain, we keep the same constraint class assignments as described in Table \ref{tab:constraint_classes}. We further use the Mardling-Aarseth hierarchical-triple stability criterion from Eq.~\eqref{eq:xoutmin_MA} with $\alpha_{\rm in}=700$ as in the constraint atlas.

\newpage
In Table \ref{tab:selected_merger_constraints} we report for each hierarchical-merger candidate the constraint classes, constraint strengths $S(\mathcal A_{\rm sec})$, retained reliable fractions $R_{\rm sec}$, and projected orbital-parameter intervals from the preliminary scans. When an inferred interval reaches a reference domain boundary, we mark these boundaries with an asterisk. 

For most of the investigated systems, the obtained constraints are weak, but a few moderate, strong, very strong and Peters-incompatible results exist. In particular, the potential hierarchical merger chain GW190514$\to$GW190521 exhibits very strong progenitor constraints, both in $a^{\rm det}_{\rm out}$ and in $e_{\rm out}$, and it even has constraints on $\mathcal M_{\rm m}$, which remains unconstrained for most tested systems. For $e_{10}<0.02$, the primary-remnant assignment satisfies the
Peters-reliability criterion with $R_{\rm sec}=1.000$, whereas the
secondary-remnant assignment has $R_{\rm sec}=0.777$ and falls below
the adopted threshold. The constraints remain strong at higher upper bounds for $e_{10}$, but $R_{\rm sec}$ quickly drops to very low values, rendering the results unreliable. Other promising systems are GW190708$\to$GW190915, GW190512$\to$GW190910, and GW170729$\to$GW170817A, which show moderate to strong constraints at low $e_{10}$, with some cases becoming Peters-incompatible for $e_{10}<0.01$.

Many of the constraints in Table \ref{tab:selected_merger_constraints} for the secular-allowed region do not change when extending the upper eccentricity limit. One possible way this can happen is that the additional target orbits fail the Peters-reliability cuts. They enlarge the full compatible region, but not its secular-compatible subset. Other possibilities are that the additional target orbits are inaccessible under the recoil and stability constraints or that they are reachable only from progenitors already allowed.

\begin{figure*}[htp!]
    \centering
    \vspace{-10pt}
    \includegraphics[width=1.0\textwidth]{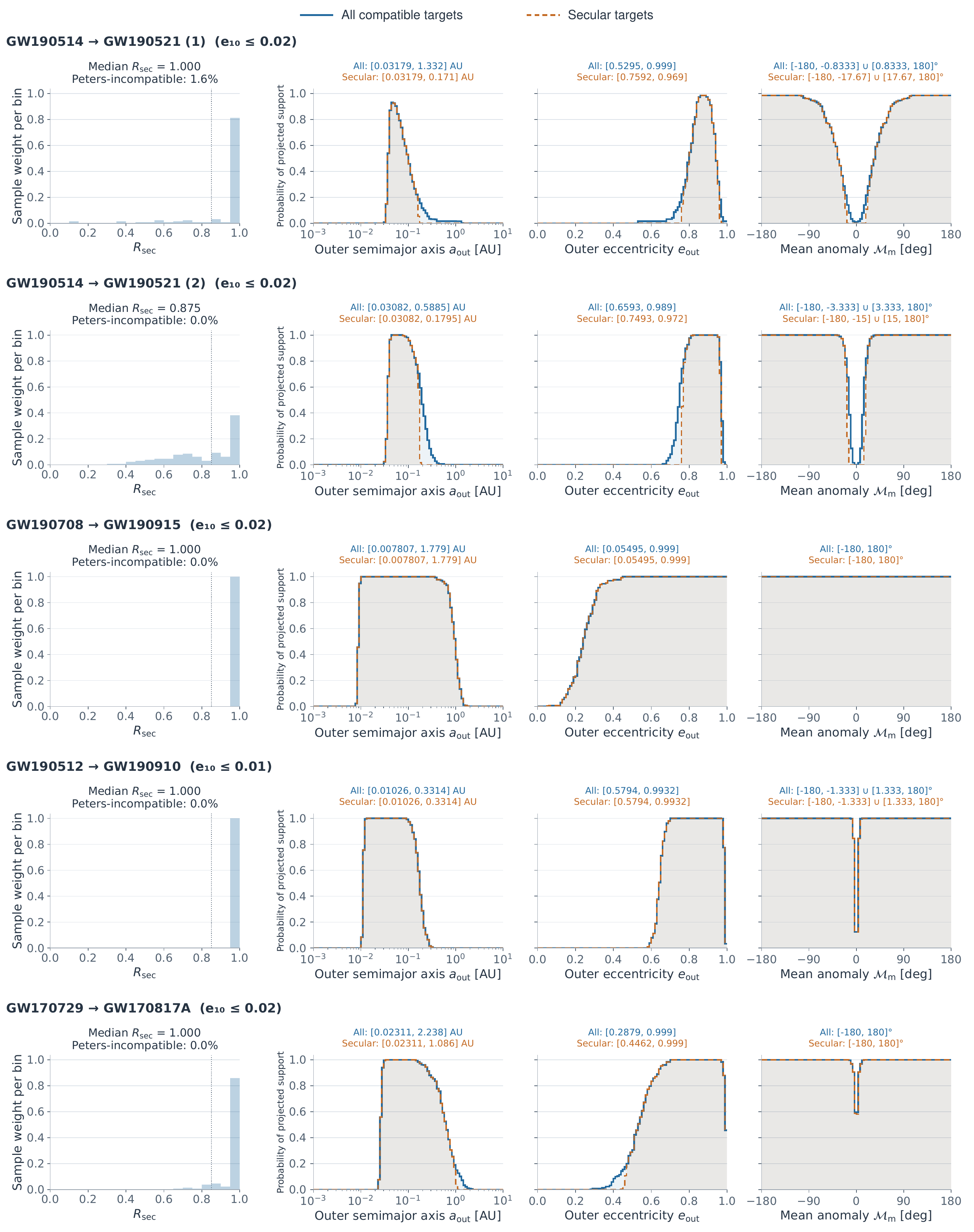}
    \caption{Posterior-weighted constraints on the progenitor outer orbit for selected hierarchical-merger candidates. From left to right, the panels show the distribution of the secular-validity fraction $R_{\rm sec}$ and the weighted projected support for the outer semi-major axis $a_{\rm out}$, eccentricity $e_{\rm out}$, and mean anomaly $\mathcal M_{\rm m}$. Blue curves include all recoil-compatible targets, orange dashed curves retain only targets satisfying the secular-validity criteria.}
    \label{fig:posterior_analyses}
\end{figure*}

To propagate measurement uncertainties into the inferred constraints, we apply our method to the full event posterior distributions for the promising systems GW190514$\to$GW190521, GW190708$\to$GW190915, GW190512$\to$GW190910, and GW170729$\to$GW170817A. We impose $e_{10}<0.02$ for all pairs except GW190512$\to$GW190910, for which we adopt $e_{10}<0.01$. For GW170729$\to$GW170817A, we could not obtain the second-event prior samples and therefore weight sample pairs solely by their mass and distance agreement. For each event pair, we draw 128 posterior sample pairs according to the matching weights and aggregate the results to obtain the distribution of $R_{\rm sec}$ and the relative frequencies with which each value of the progenitor outer-orbit parameters is allowed.

The results of the posterior analyses are shown in Fig.~\ref{fig:posterior_analyses}. Above each panel, we give the median $R_{\rm sec}$ or the union of the allowed intervals across all evaluated posterior draws. The orbital-support curves show the weighted fraction of draws that admit each parameter range. Values close to unity indicate ranges supported by most draws, while low-amplitude tails indicate ranges supported by only a few. These curves are not posterior probability densities for the outer-orbit parameters, and the reported unions are not credible intervals.

Most draws lie in the Peters-reliable regime. However, both remnant assignments for GW190514$\to$GW190521 as well as GW170729$\to$GW170817A include draws with lower $R_{\rm sec}$, especially the secondary-remnant assignment. The blue curves include all compatible targets, whereas the orange dashed curves include only targets satisfying the secular-validity criteria. Overall, informative constraints on the progenitor outer orbit remain after propagating the measurement uncertainties.

\section{Summary and discussion}
\label{sec:summary}

In this work, we presented a novel method for constraining the progenitor outer orbits of sequential black-hole mergers in hierarchical triples. Given an adopted upper bound on the first-merger recoil velocity, the observational inputs are the black-hole masses, the inter-merger delay, and the eccentricity of the second binary at a reference gravitational-wave frequency. We use the Peters equations to construct post-first-merger target orbits consistent with these inputs and determine the minimum recoil velocity required to reach them from each progenitor configuration. Imposing the recoil ceiling and a hierarchical-stability criterion then constrains the outer semi-major axis, eccentricity, and orbital phase at the first merger. The full procedure is summarized in Fig.~\ref{fig:method_chart}.

The entire method can be evaluated consistently in either the detector frame or the source frame. Working in the detector frame allows the measured masses, inter-merger delay, and reference frequency to be used directly, without requiring a redshift estimate. A common source redshift is only needed to convert the inferred detector-frame separations into source-frame physical lengths, through $a_{\rm out}^{\rm src}=a_{\rm out}^{\rm det}/(1+z)$. The resulting constraints remain conditional on the assumed sequential-merger association and the adopted dynamical model. They do not independently establish that two observed events share a common progenitor.

Our constraint atlas demonstrates that informative outer-orbit constraints can be obtained across a range of masses, merger delays, and detector-band eccentricities. The strongest constraints occupy a narrow band in the intrinsic parameter space, whose location is described by the empirical relation in Eq.~\eqref{eq:strong_constraint_fit}. The constraints for the three reconstructed coordinates behave very differently: the semi-major axis remains constrained over a comparatively broad region, tight eccentricity constraints are concentrated more closely around the strong-constraint band, and the orbital phase is generally the least constrained quantity. A small allowed progenitor volume therefore does not imply equally restrictive bounds on every orbital parameter.

The typical inter-merger delays for which strong constraints can be obtained generally decrease as the detector-band eccentricity increases. At $f_{\rm ref}=10$ Hz, nearly circular second mergers can yield strong constraints for delays of several months to a few years. Eccentricities around $0.05$ shift the characteristic delays to weeks or months, while eccentricities above approximately $0.1$ can yield strong constraints at delays of order days. Among systems satisfying our Peters-reliability criterion, the aggregate trends favor stronger constraints at shorter delays and higher detector-band eccentricities. Strong, reliable constraints nevertheless occur across a broad range of the sampled system parameters, although no Peters-reliable systems were found at the shortest sampled inter-merger delays.

Assessing the applicability of the Peters approximation is central to our method. We distinguish target orbits satisfying our applicability criteria \eqref{eq:peters_validity_cuts} from those for which orbit-averaged evolution is less well justified. For each system, the retained fraction $R_{\rm sec}$ measures how much of the Peters-inferred compatible progenitor region remains supported after imposing these cuts, with $R_{\rm sec}\geq0.85$ defining our operational Peters-reliability criterion. This diagnostic is neither a probability that the inferred constraints are correct nor an error estimate for their boundaries. In particular, even $R_{\rm sec}=1$ does not establish that a more accurate dynamical treatment would admit no additional solutions outside the Peters-inferred region. Likewise, Peters-incompatibility denotes the absence of compatible solutions within the adopted approximation and search domain, rather than a definitive exclusion of a hierarchical-triple origin. Nevertheless, we expect the results for Peters-reliable systems, including the transition to Peters-incompatibility as their recoil-compatible regions shrink to zero, to provide useful first approximations to those obtained with a more accurate dynamical treatment.

For a quick assessment of whether a system is likely Peters-reliable, one can use our derived expression for the approximate critical detector-band eccentricity in Eq.~\eqref{eq:atlas_critical_eccentricity}. On our sampled parameter domain, the highest fraction of Peters-reliable systems is achieved for systems with low total masses, low detector-band eccentricities and intermediate inter-merger delays. The mass ratio of the post-first-merger binary has only a weak effect on both the constraint strength and the Peters-reliability.

Comparisons with post-Newtonian integrations support the usefulness of the approximation within the tested regime while illustrating its limitations. For the two diagnostic categories inside the applicability cuts, the median absolute merger-time discrepancies are approximately $4.7\%$ and $7.2\%$, with median absolute eccentricity discrepancies of $8.3\times10^{-4}$ and $2.1\times10^{-3}$ at 10 Hz. Much larger discrepancies and stronger initial-phase dependence occur when the time to merger becomes comparable to or shorter than the initial orbital period.

We further demonstrated that applying our method to selected hierarchical-merger candidates from the literature can yield informative constraints on the progenitor outer orbit, even after accounting for gravitational-wave measurement uncertainties. The posterior-weighted support curves distinguish parameter ranges compatible with most sampled systems from low-support extensions admitted by only a few draws. Most draws satisfy the adopted reliability criterion, but GW170729$\to$GW170817A as well as both remnant assignments for GW190514$\to$GW190521 include draws with substantially lower $R_{\rm sec}$, particularly the secondary-remnant assignment.
\\
\\
The presented examples highlight that precise constraints on the individual system parameters are indispensable for accurate constraints on the progenitor orbital parameters. In particular, the constraint atlas and the results in Table \ref{tab:selected_merger_constraints} illustrate how strongly the constraint strength and Peters-reliability depend on the upper limits on the eccentricity. Future gravitational-wave observatories will likely constrain eccentricity more accurately, particularly through improved low-frequency sensitivity. However, distinguishing eccentricity from spin precession remains challenging, especially for short, high-mass signals \citep{Buscicchio2021, XuHamilton2023, Saini2024}. For accurate constraints on the semi-major axis in the source frame, one additionally needs strong constraints on the source redshifts. Future detectors should enable more precise luminosity-distance measurements, but the inferred redshifts will remain subject to uncertainties from the distance--inclination degeneracy, weak gravitational lensing, and the assumed cosmological model \citep{Pieroni2022, BorhanianSathyaprakash2024}. Precise measurements of the masses, spins, redshifts, and sky locations are essential both for constraining progenitor orbits and for assessing whether individual events belong to the same hierarchical-merger chain.

Another weakly constrained quantity that our method heavily relies on is the recoil velocity after the first merger. In our examples, we adopt a fiducial value of $v_{\rm k,max}=500~{\rm km\,s^{-1}}$, motivated by numerical-relativity estimates of the maximum recoil for non-precessing black-hole binaries (e.g.~Ref.~\cite{Healy2014}). However, generic precessing configurations with large in-plane spins can produce recoils approaching $5000~{\rm km\,s^{-1}}$ \citep{Lousto2011}. Future analyses could replace a fixed ceiling with event-specific recoil information inferred from binary masses and spins using numerical-relativity remnant models \citep{Varma2020, Ranjan2025}. This would also allow correlations between remnant properties and recoil to be treated consistently. Direct measurements of the recoil's Doppler imprint may provide complementary information, although they primarily constrain its line-of-sight component \citep{GerosaMoore2016, Mahapatra2024}.

Future gravitational-wave observations may also constrain the outer orbits of hierarchical triples directly through modulations of the inner binary's signal. The motion of the binary around the triple's center of mass produces Doppler shifts and variations in the signal arrival time that encode information about the tertiary companion and its orbit \citep{Meiron2017,Wong2019}. Long-duration observations with space-based detectors such as LISA are particularly promising: when the outer orbital period is comparable to or shorter than the observation time, the modulation may allow measurements of the outer period and eccentricity, whereas shorter observations may primarily constrain the line-of-sight acceleration \citep{Robson2018}. In general, the attainable precision significantly depends on the system parameters and viewing geometry. Such measurements would complement our reconstruction from successive mergers by providing independent orbital information that could further constrain the hierarchical-triple progenitor configuration.
\\
\\
In this study, we applied our method to hierarchical systems consisting entirely of black holes. Nevertheless, the orbital reconstruction should be generalizable to triples consisting of other types of compact objects, provided that their mergers supply the required inputs and the relevant dynamics remain adequately described by the adopted model. For some compact-object systems, finite-size effects such as tides, mass transfer, collisions, and disruption can invalidate the point-mass and gravitational-wave-driven treatment. Additionally, asymmetric ejecta and matter interactions during the merger may require a more detailed treatment than instantaneous mass loss and recoil.

A natural extension of our method is to incorporate environmental effects, such as gas damping, and to replace the Peters evolution with a more accurate description of relativistic and nonadiabatic dynamics. Although repeated direct numerical integrations would make the required inverse search computationally demanding, surrogate models mapping the relevant initial conditions to the merger time and reference eccentricity could make this approach practical. Our framework and the accompanying \texttt{triple-check} code provide a foundation for such extensions and help identify systems that would benefit most from more detailed dynamical modeling.

\begin{figure*}[htp!]
    \centering
    \vspace{-10pt}
    \includegraphics[width=1.0\textwidth]{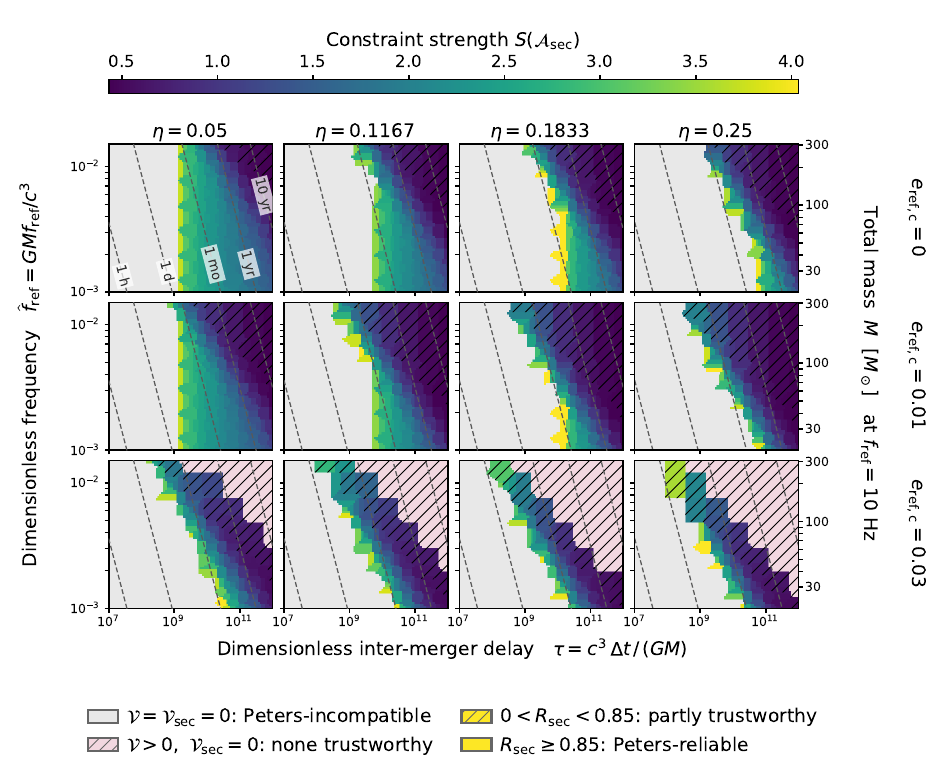}
    \vspace{-20pt}
    \caption{Constraint strengths for the mass ordering branch $m_3>m_{\rm rem}$ and their dependence on the dimensionless frequency $\widehat f_{\rm ref}$ (or total mass), inter-merger delay $\tau$, symmetric mass ratio $\eta$, and eccentricity at the reference frequency $e_{\rm ref,c}$. This plot corresponds to the one in Fig.~\ref{fig:atlas_S} for $m_{\rm rem}>m_3$, but here only the first three eccentricity values are shown.}
    \label{fig:atlas_S_m3_gt_mrem}
\end{figure*}

\section*{Acknowledgments}
FMH has been supported by the Deutsche Forschungsgemeinschaft (DFG) under Grant No. 406116891 within the Research Training Group RTG 2522/1. FMH also acknowledges support from the German Academic Exchange Service (DAAD) through a doctoral research scholarship.

\textsc{GPT-6 Astra} (OpenAI) was used to assist with code development and the drafting and editing of this manuscript. The authors reviewed and verified the resulting code and take full responsibility for the analyses, results, and final content of the paper.

\section*{Data Availability}
The code \textsc{triple-check}, which was used for performing the calculations and generating some of the figures in this manuscript, is publicly available \citep{triple-check}. All raw data corresponding to the findings in this manuscript are not publicly available upon publication because it is not technically feasible and/or the cost of preparing, depositing, and hosting the data would be prohibitive within the terms of this research project. The data are available from the authors upon reasonable request.

\appendix
\section{Constraint atlas for \texorpdfstring{$m_3>m_{\rm rem}$}{m3>mrem}}
\label{sec:appendix}
In this appendix, we discuss selected differences between the constraint atlas for the mass-ordering branch $m_3>m_{\rm rem}$ and the other branch with $m_{\rm rem}>m_3$. Rather than reproducing the full set of atlas figures, we focus on the qualitative differences at low eccentricities. The systematic shift in the location of the diagonal strong-constraint band is already described by Eq.~\eqref{eq:strong_constraint_fit}.

Fig.~\ref{fig:atlas_S_m3_gt_mrem} shows the first three eccentricity rows of the constraint strength for the mass-ordering branch $m_3>m_{\rm rem}$. The clearest qualitative differences from the other branch occur in the top two rows, where nearly vertical bands appear at small symmetric mass ratios and low values of $e_{\rm ref,c}$. At fixed $\tau$, $\eta$, and mass ordering, varying $\widehat f_{\rm ref}$ changes the frequency at which the post-merger eccentricity is evaluated, while leaving the merger-time, recoil, and progenitor-stability conditions unchanged. The detector eccentricity windows in these two rows include zero and therefore impose only an upper limit on the eccentricity. Along the vertical bands, the allowed post-merger orbits have already circularized sufficiently to satisfy this upper limit at the lowest reference frequencies shown. Increasing $\widehat f_{\rm ref}$ reduces their eccentricities further, so the same configurations continue to satisfy the eccentricity condition and remain allowed. Additional post-merger targets can then satisfy this condition, but they do not enlarge the allowed progenitor set over the frequency range of the plateau: they are either inaccessible under the remaining constraints or reachable only from progenitor configurations already included in that set. Consequently, the allowed volume and the corresponding constraint strength remain approximately unchanged across this frequency range. The existence and extent of these bands depend on the adopted inner-separation and stability prescriptions, as well as on the choice of the lower separation bound $x_{\rm out,min}$. Under our prescription, at fixed total mass and $\eta$, assigning the lighter component to the remnant gives a smaller pre-merger inner binary and consequently allows stable progenitors with tighter outer orbits. These configurations can produce the required inter-merger delays while simultaneously satisfying the recoil and detector-eccentricity constraints. For example, at $e_{\rm ref,c}=0$, $\eta\simeq0.1167$, and $\tau\simeq1.33\times10^{10}$, the recovered solutions in the frequency plateau occupy the separation range $x_{\rm out}\simeq503$--$587$. In the opposite mass-ordering branch, by contrast, the adopted stability criterion requires $x_{\rm out}\gtrsim1263$ even for a circular orbit.

\bibliography{apssamp} 

\end{document}